\documentclass[pre,aps,superscriptaddress,reprint]{revtex4-1}
\usepackage{graphicx,rotating,subfigure}
\usepackage{float}
\usepackage{amsmath}
\usepackage{pstricks}
\usepackage{gensymb}

\def\Conc01{/home/damien/Mesoplasticity/Composite/N64/epsi10/bias01}

\begin{document}

\title{Finite size effects in a model for plasticity of amorphous composites}

\author{Botond Tyukodi}
\affiliation{
Laboratoire PMMH,
CNRS-UMR 7636/ESPCI/UPMC/Univ. Paris 7 Diderot,
10, rue Vauquelin, 75231 Paris cedex 05, France $\mbox{}$
}
\affiliation{Babe\c{s}-Bolyai University, Physics department, 
RO-3400, Cluj, Romania $\mbox{}$
}

\author{Claire A. Lemarchand}
\email{clairel@ruc.dk}
\affiliation{
  DNRF Centre ``Glass and Time'', IMFUFA, Department of Sciences, 
  Roskilde University, Postbox 260, DK-4000 Roskilde, Denmark $\mbox{}$
}

\author{Jesper S. Hansen}
\affiliation{
  DNRF Centre ``Glass and Time'', IMFUFA, Department of Sciences, 
  Roskilde University, Postbox 260, DK-4000 Roskilde, Denmark $\mbox{}$
}

\author{Damien Vandembroucq}
\affiliation{ 
Laboratoire PMMH,
CNRS-UMR 7636/ESPCI/UPMC/Univ. Paris 7 Diderot,
10, rue Vauquelin, 75231 Paris cedex 05, France $\mbox{}$
}

\begin{abstract}
We discuss the plastic behavior of an amorphous matrix reinforced by
hard particles. A mesoscopic depinning-like model accounting for
Eshelby elastic interactions is implemented. 
Only the effect of a plastic disorder is considered.
Numerical results show a
complex size-dependence of the effective flow stress of the amorphous
composite. In particular the departure from the mixing law shows
opposite trends associated to the competing effects of the matrix and
the reinforcing particles respectively.  The reinforcing mechanisms
and their effects on localization are discussed. Plastic strain is
shown to gradually concentrate on the weakest band of the system. This
correlation of the plastic behavior with the material structure is
used to design a simple analytical model. The latter nicely captures
reinforcement size effects in $-(\log N/N)^{1/2}$ observed
numerically. Predictions of the effective flow stress accounting for
further logarithmic corrections show a very good agreement with
numerical results.
\end{abstract}

\date{\today}
\maketitle

\noindent Corresponding author: C. A. Lemarchand, E-mail: clairel@ruc.dk 

\section{Introduction\label{Intro}}

The introduction of foreign particles in an amorphous matrix has for
  long been used for the strength reinforcement of disordered
  materials~\cite{Torquato-book02}. The most classical strategy
  consists in adding rigid particles or fibers in order to enhance the
  elastic properties of the composite material. 

An additional or alternative strategy consists in the modification of
the plastic properties. Here, the effect on the overall strength is
more delicate. The introduction of hard particles in a very ductile
matrix tends to increase the effective yield stress, hence the
strength. A good illustration of this approach can be found in the
development of materials for road
pavements~\cite{Branthaver-SHRP93,Anderson-SHRP94,Chen-JMCE98,You-CBM12}:
mineral micrometer scale fillers are introduced in a viscous bitumen
to make it viscoplastic; the obtained mastic asphalt is then
reinforced through the addition of millimetric to centimetric
aggregates. More recently the introduction of a ductile phase has been
used to reinforce metallic
glasses~\cite{Hofmann-Nat08,Ferry-MRS13}. In this case, the ductility of the
second phase enables one to control the development of shear-bands,
thus preventing the nucleation of cracks. A reinforcement effect is
obtained despite the fact that the effective yield stress of the
amorphous composite is lowered with respect to that of the
matrix.

The understanding of the plastic behavior of amorphous composites thus
appears to be crucial in the design of modern materials. Efforts
in theoretical and numerical modelling have been recently performed to
study the effects of micro-alloying in metallic
glasses~\cite{Samwer-APL13,Samwer-ActaMat14} and of the addition of
aggregates in asphalt mixtures~\cite{Al-Rub09,Al-Rub11,Al-Rub12}.

From the theoretical mechanics point of view, the determination of effective
mechanical properties is a matter of homogenization. While this field
has been intensively explored in case of elastic
properties~\cite{Torquato-book02}, results are much more scarce for
non-linear behaviors like fracture~\cite{RVH-EJMA03,PVR-PRL13} or
plasticity~\cite{Ponte-Castaneda-PRSA92,Debotton-IJSS95,Willis-JMPS04,Turgeman-MRC11,Suquet-IUTAM14}. In
particular, standard homogenization approaches fail to account for
size effects~\cite{Ponte-Castaneda-PRSA92,Debotton-IJSS95}. Only the
development of strain-gradient theories (relying on the introduction
of an ad-hoc internal length scale) has so far succeeded to reproduce
size dependence~\cite{Willis-JMPS04}. Still, these approaches only
predict the mean behavior and cannot cope with sample-to-sample
fluctuations.

Here we develop an alternative approach, based on the recent
development of depinning-like mesoscopic models of amorphous
plasticity~\cite{BulatovArgon94a,BVR-PRL02,Picard-PRE02,Schuh-ActaMat09,TPVR-CRM12,Nicolas-SM14}. The
modelling of amorphous plasticity and rheology of complex fluids has
seen much progress in recent years~\cite{RTV-MSMSE11} and a family of
mesoscopic models~\cite{BVR-PRL02,TPVR-CRM12} has emerged that rely on
two main ingredients: local plastic thresholds (amorphous plasticity
results from series of local rearrangements of the amorphous
structure~\cite{Argon-ActaMet79,FalkLanger-PRE98}) and account of
elastic interactions (local plastic events occur in a surrounding
elastic matrix and induce internal stresses~\cite{Eshelby57}). These
models show scaling properties close to the effective yield stress
(here seen as a critical threshold) and thus exhibit statistical size
effects. Another nice feature of these models is their ability to
reproduce localization and shear-banding
behaviors~\cite{VR-PRB11,Homer-ActaMat14}. The effect of crystalline
inclusions in an amorphous matrix has recently been discussed along
such lines in Ref.~\cite{Homer-ActaMat15} with a particular emphasis
on the localization behavior.

Here we specialize the model recently presented
in~\cite{TPVR-PRE11,TPVR-CRM12} to the case of amorphous composites by
considering a binomial distribution of local plastic stress thresholds
in order to reproduce the introduction of a fraction of hard particles
in an amorphous matrix. The simplistic model presented in the
following will obviously not be able to give a realistic account of
the whole richness of the mechanical behavior of amorphous
composites. However results are expected to be generic for this class
of materials.

In section~\ref{Model} we introduce the model, in
section~\ref{Size-effects} we present the complex size dependence of
the yield strength measured on the amorphous matrix and amorphous
composites with a growing fraction of particles. In
section~\ref{Hardening}, we discuss the hardening mechanisms at play
in amorphous composites and the localization behavior. We emphasize in
particular the interplay between the gradual localization of the
plastic deformation and the building of a strongly correlated internal
stress field. Elaborating on the numerical observations, we present
in section~\ref{Analytical-Model} an analytical model that accounts
quantitatively for the size effects of the effective yield strength of
amorphous composites. Mathematical details of the model are provided
in a separated appendix. Our main findings are finally summarized in
section~\ref{Conclusion}.

\section{Modelling amorphous plasticity: from glasses to amorphous composites
\label{Model}}

The modelling of amorphous plasticity has recently given rise to an
increasing interest~\cite{RTV-MSMSE11}. Unlike crystalline plasticity
that results from the motion of dislocations of the ordered lattice,
amorphous plasticity results from series of localized rearrangements
of the disordered
structure~\cite{Spaepen-ActaMet77,Argon-ActaMet79,FalkLanger-PRE98}. Such
local plastic events induce internal stresses within the surrounding
material~\cite{Maloney-PRL04b,Tanguy-EPJE06}. The latter can be seen as
an elastic matrix around a plastic inclusion and the stress associated
to the rearrangement computed in the spirit of the problem of the
eigen strain early introduced by Eshelby~\cite{Eshelby57}.

\begin{figure}
\includegraphics[width=0.48\columnwidth]{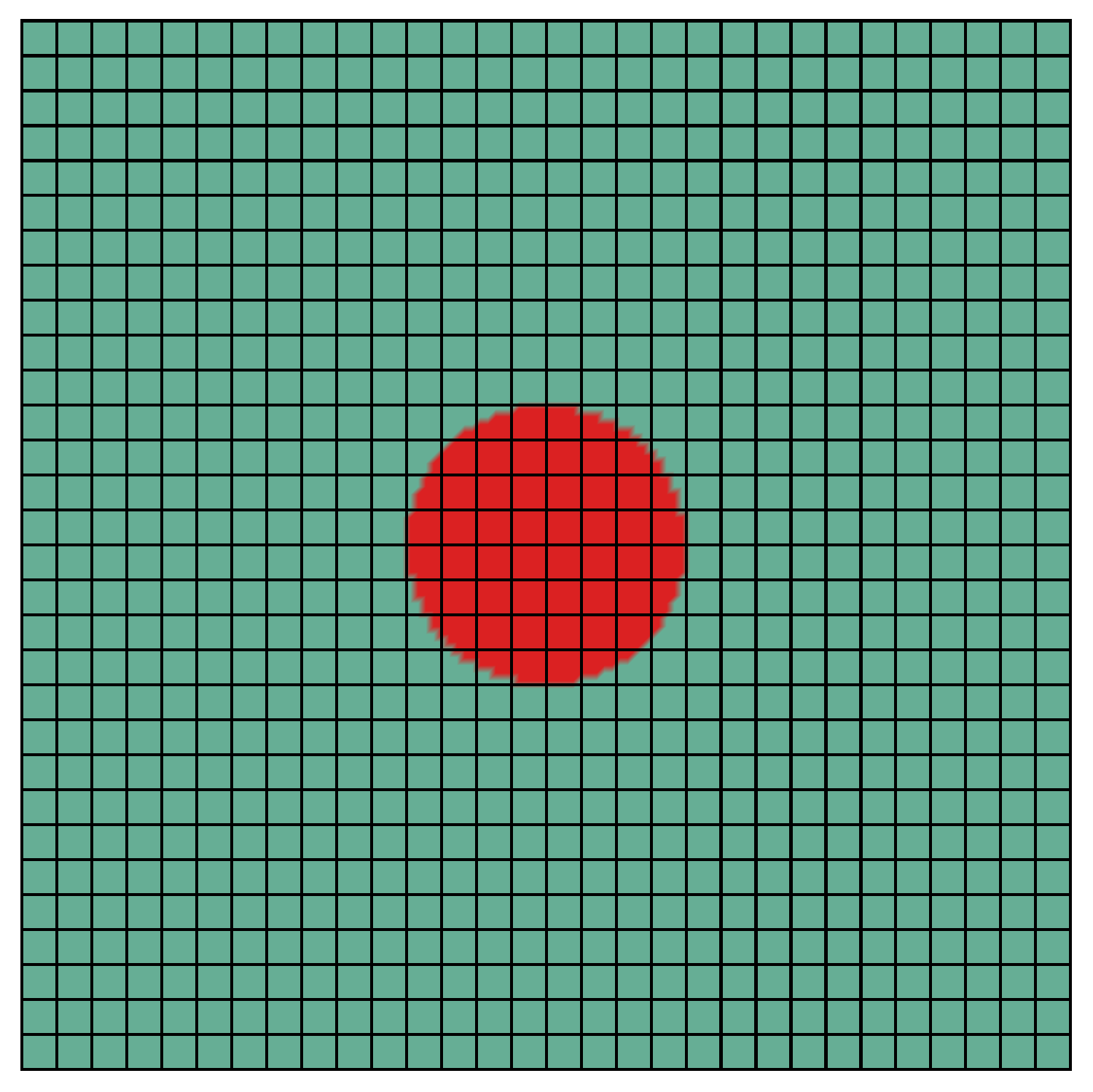}
\includegraphics[width=0.48\columnwidth]{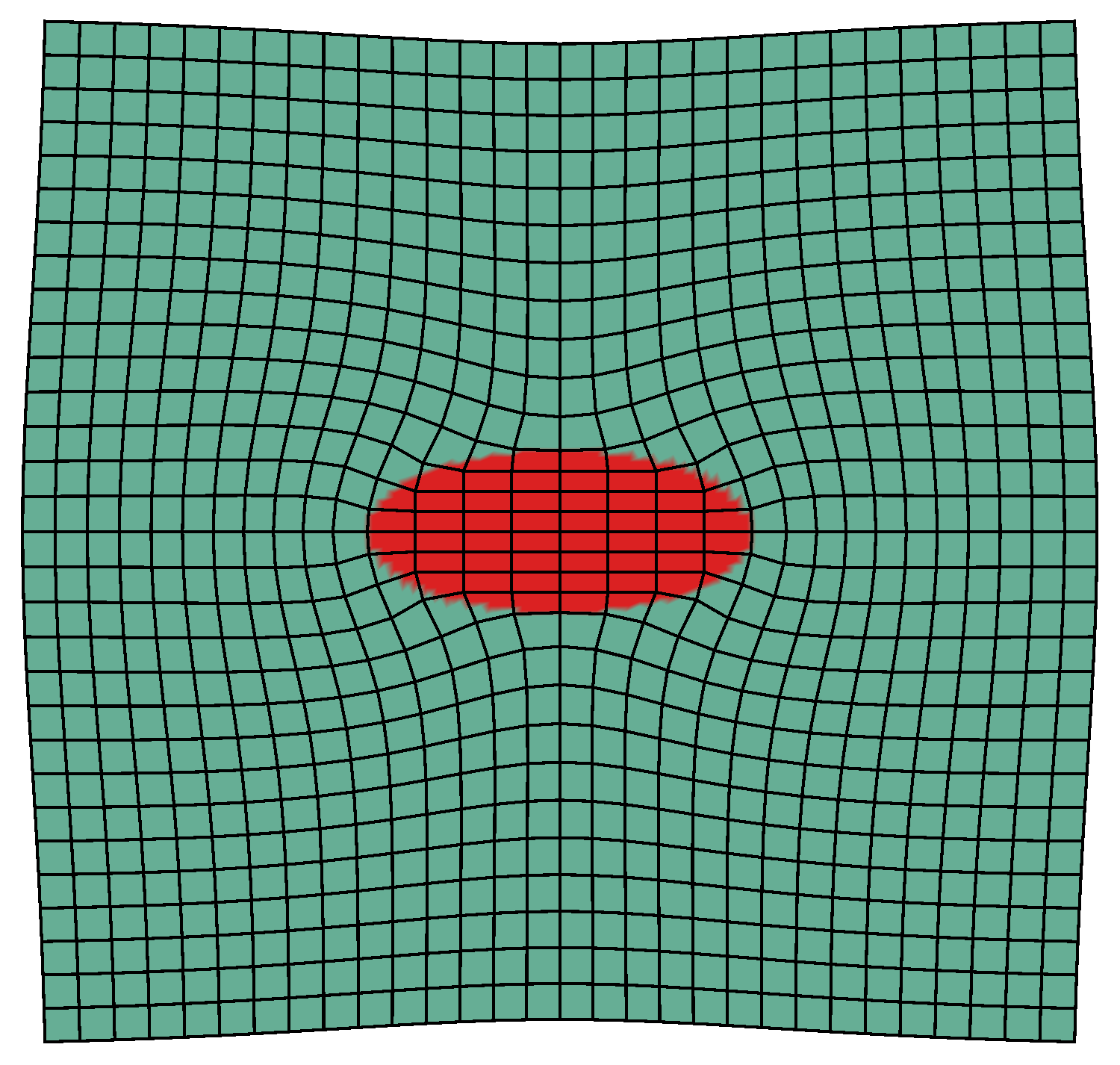}
\includegraphics[width=0.48\columnwidth]{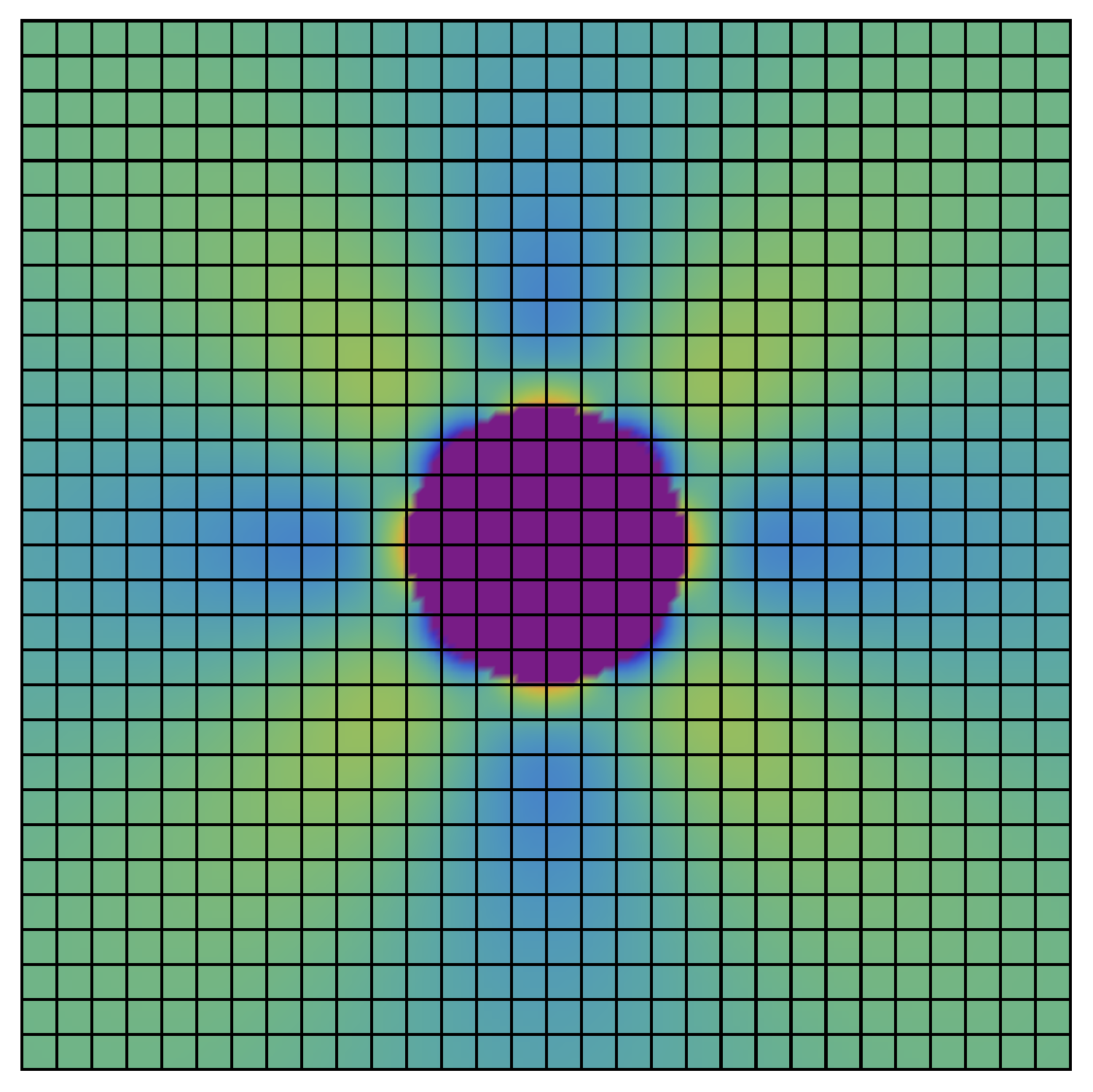}
\includegraphics[width=0.48\columnwidth]{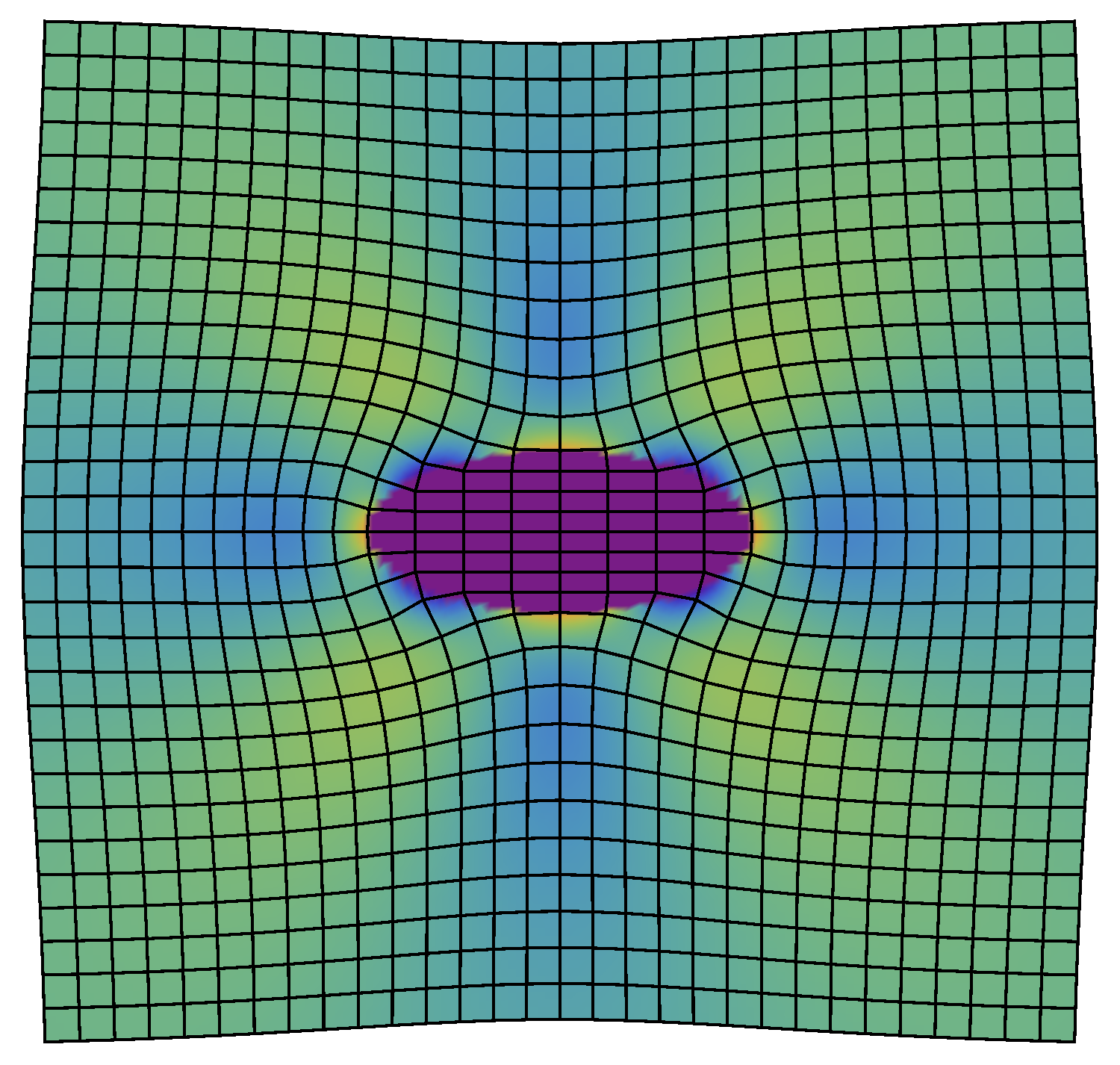}
\includegraphics[width=0.48\columnwidth]{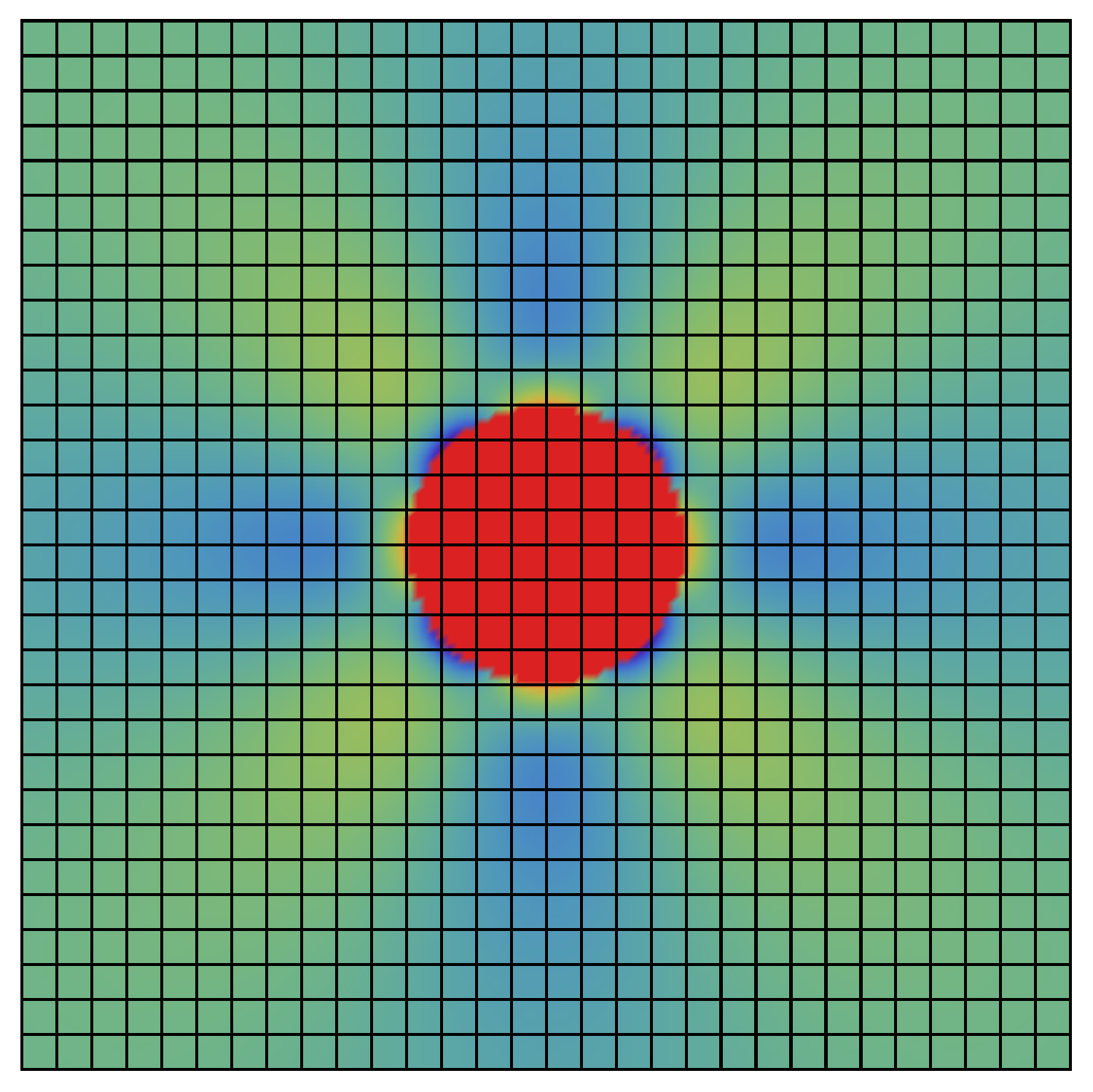}
\includegraphics[width=0.48\columnwidth]{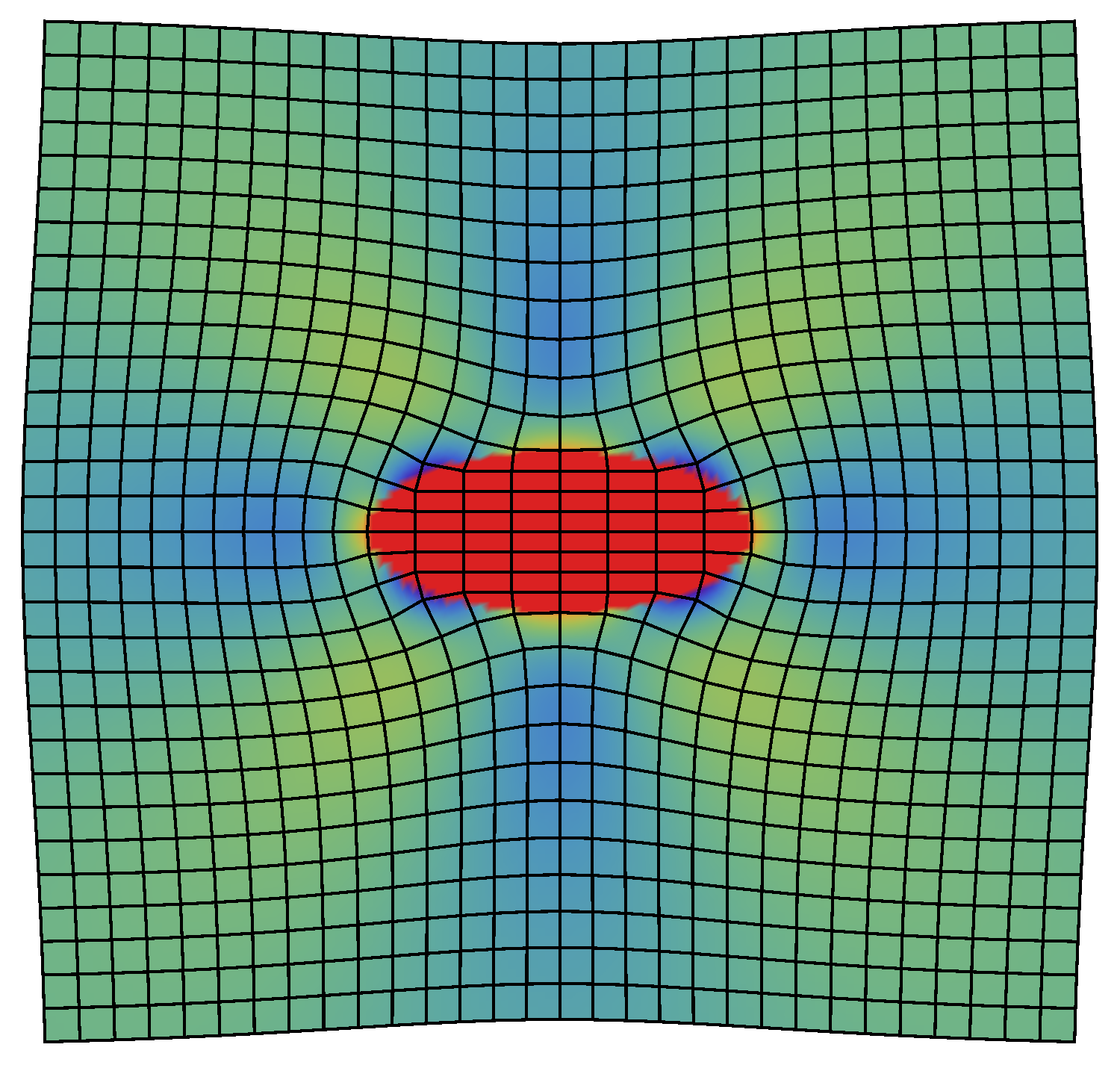}
\includegraphics[width=0.04\columnwidth,angle=-90]{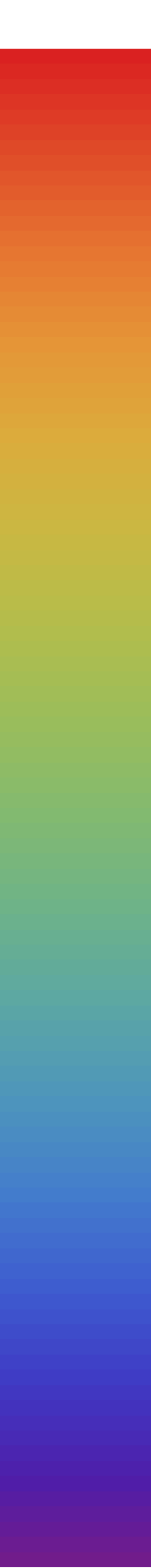}
\caption{Effect of a plastic inclusion in an elastic matrix submitted
  to a pure shear biaxial loading
  $\Sigma=\Sigma_{yy}=-\Sigma_{xx}$. Only the spherical inclusion
  experiences plasticity. The first row represents the plastic strain
  $\varepsilon^{pl}$, the second row the elastic strain
  $\varepsilon^{el}$ and the third row the total strain
  $\varepsilon=\varepsilon^{el}+\varepsilon^{pl}$. The strain fields
  are represented by a color scale on the reference mesh on the left
  column and on a deformed mesh on the right column. One recognizes
  the traditional quadrupolar symmetry associated to the Eshelby
  inclusion.}
\label{fig:Eshelby}
\end{figure}

In Fig.~\ref{fig:Eshelby} we show the effect of such a plastic inclusion
on the surrounding matrix. The inclusion has experienced a pure shear
$\varepsilon^{pl}_{xx}=-\varepsilon^{\mathrm{pl}}_{yy}=\varepsilon^{\mathrm{pl}}_0$. The
amplitude of the shear strain $\varepsilon=\varepsilon_{xx} -
\varepsilon_{yy}$ is represented by a color scale on a reference
undeformed grid (left column) and on a grid deformed according to the
total displacement (right column). The first row shows the plastic
strain $\varepsilon^{\mathrm{pl}}$, non-zero only within the inclusion. The
second row shows the elastic strain field
$\varepsilon^{\mathrm{el}}$. The latter is negative within the
inclusion. Outside the inclusion it exhibits a quadrupolar symmetry:
negative along the axes at $0$ and $90$ degrees, and positive along
the directions at $\pm45$ degrees. The third row shows the total
strain
$\varepsilon=\varepsilon^{\mathrm{pl}}+\varepsilon^{\mathrm{el}}$.

The precise expression of the internal stress field
$\sigma^{\mathrm{el}}$ depends on the details of the plastic
strain field and the geometry of the rearranging region but in the
far field, the dominant term obeys the universal form:
\begin{equation}
\sigma^{\mathrm{el}}=\mu \varepsilon^{\mathrm{pl}}_0 \mathcal{A} \frac{\cos(4\theta)}{r^2}\;,
\label{quadrupolar-stress}
\end{equation}
where $\varepsilon^{\mathrm{pl}}_0$ and $\mathcal{A}$ are the mean plastic strain
experienced by the inclusion and the area of the inclusion, respectively.

The quadrupolar elastic interaction associated with localized plastic
events is the essential ingredient of the recent models of amorphous
plasticity and/or rheology of complex fluids. Indeed its anisotropic
nature is responsible for non-trivial
behaviors~\cite{TPVR-CRM12,Wyart-PNAS14}. In particular the presence
of multiple soft modes of the elastic interaction, corresponding to
the existence of extended modes of plastic strain which do not induce
internal stresses, e.g., shear bands, is responsible for a complex
localization behavior.

\subsection{A mesoscopic model of amorphous plasticity}

Following the model introduced in Refs.~\cite{BVR-PRL02, TPVR-CRM12},
the system is discretized on a two-dimensional square lattice with a
lattice constant that is larger than the typical size of the plastic
reorganizations. Each site $(i, j)$ has an internal stress
$\sigma^{\mathrm{el}}_{ij}$,
a local plastic threshold $\sigma^{\mathrm{c}}_{ij}$, and
a local plastic strain $\varepsilon^{\mathrm{pl}}_{ij}$.
A pure shear external loading is considered:
$\Sigma_{xx}^{\mathrm{ext}}=-\Sigma_{yy}^{\mathrm{ext}}=\Sigma^{\mathrm{ext}}$. It
is assumed that the reorganizations at a microscopic scale obey the
same symmetry as the external loading, i.e., a site $(i, j)$ undergoes a plastic
deformation in pure shear:
$\varepsilon_{xx}^{\mathrm{pl}}=-\varepsilon_{yy}^{\mathrm{pl}}=\varepsilon^{\mathrm{pl}}_{ij}$. A
local criterion of plasticity is considered, the limit of elasticity
is thus defined for a site $(i, j)$ by:
\begin{equation}
\Sigma^{\mathrm{ext}} + \sigma^{\mathrm{el}}_{ij} \le \sigma^{\mathrm{c}}_{ij}\;.
\label{criterion}
\end{equation}
Values of
$\sigma^{\mathrm{c}}$ are drawn from a random distribution. No spatial
correlations are considered.

Whenever the criterion is locally satisfied, say on site $(i_0,j_0)$
the site undergoes an incremental plastic strain $\delta
\varepsilon_0^{\mathrm{pl}}$. This value is drawn from a uniform
distribution in $[0,d_0]$. To account for the structural change
experienced by the rearranging zone, the local plastic threshold is
updated to a new value.  As discussed above, the local plastic event also
induces an incremental internal stress on every lattice site $(i, j)$:
\begin{equation}
\delta\sigma^{\mathrm{el}}_{ij} = G^{\mathrm{el}} * \delta \varepsilon_0^{\mathrm{pl}}
\;,
\label{internal-stress}
\end{equation}
where the star denotes the convolution operation and $G^{\mathrm{el}}$ is a
quadrupolar kernel accounting for the elastic reaction of the matrix
to a unit plastic event. Here we consider bi-periodic boundary
conditions and $G^{\mathrm{el}}$ is computed from Fourier
space~\cite{TPVR-CRM12,Budrikis-PRE13,TPRV-preprint15}. 

The system is driven with an extremal dynamics: only one site is
deformed per iteration step. An iteration step corresponds to $(i)$
identify the weakest site $(i_0,j_0)$ for a given configuration, $(ii)$ update the
plastic strain $\varepsilon^{\mathrm{pl}}_{i_0,j_0}$ and the plastic threshold
$\sigma^{\mathrm{c}}_{i_0,j_0}$ at this particular site and $(iii)$ update the
internal stress $\sigma^{\mathrm{el}}$ all over the system. A new
configuration is thus obtained and the next iteration can be
performed. 
Extremal dynamics~\cite{BVR-PRL02} is a way of driving the
system at a vanishing strain rate, very close in spirit to the
athermal quasi-static driving used in some atomistic
simulations~\cite{Maloney-PRL04a,Maloney-PRE06}. 
Note that the same
model can be driven with other kinds of dynamics, e.g.,
constant stress, kinetic Monte Carlo.

A direct outcome of a simulation is the evolution of the external
stress $\Sigma^{\mathrm{ext}}$ versus the average plastic strain
$\langle \varepsilon^{\mathrm{pl}} \rangle$, where the average
$\langle \cdot \rangle$ represents the average over the different
sites at a particular iteration step. The average plastic strain
$\langle \varepsilon^{\mathrm{pl}} \rangle$ is directly proportional
to the number of iteration steps so that $\langle
\varepsilon^{\mathrm{pl}} \rangle$ can be seen as a fictitious time.

\begin{figure}
\begin{center}
\includegraphics[width=0.95\columnwidth]{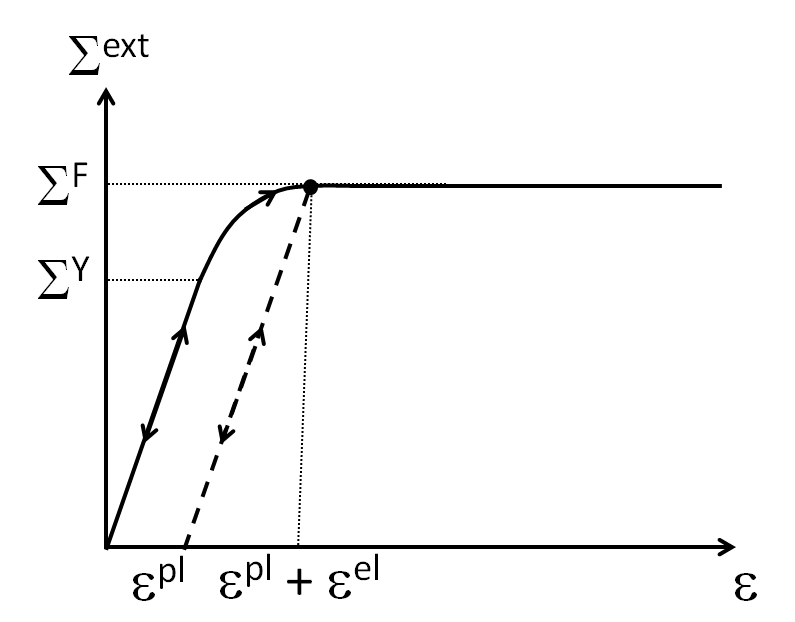}
\caption{\label{fig:stressStrainTypical}Sketch of a simple plastic
  behavior. Plasticity sets in at yield stress $\Sigma^{\mathrm{Y}}$,
  a hardening stage follows until a stress plateau is reached. The
  latter stress value defines the flow stress
  $\Sigma^{\mathrm{F}}$. The plastic strain
  $\varepsilon^{\mathrm{pl}}$ is defined as the total strain
  $\varepsilon$ minus the elastic strain $\varepsilon^{\mathrm{el}}$.}
\end{center}
\end{figure}

In Fig.~\ref{fig:stressStrainTypical} we give a sketch of a simple
plastic behavior. A typical stress-strain curve obtained upon
monotonous loading is shown. A (reversible) elastic behavior is first
observed up to the yield stress value $\Sigma^{\mathrm{Y}}$. Above
this value, plasticity sets in (a residual plastic strain is obtained
upon unloading). The following curvature of the stress-strain curve is
characteristic of a hardening behavior: if a unloading/loading cycle
is performed, a new (larger) value of the elastic limit is obtained. A
stress plateau is eventually reached that defines the ultimate flow
stress $\Sigma^{\mathrm{F}}$.

In the present framework, the external loading is not
monotonous. Rather, the external stress $\Sigma^{\mathrm{ext}}$ is a
fluctuating quantity which is adapted at each iteration step so that
the criterion of the weakest site is satisfied.  The macroscopic flow
stress $\Sigma^{\mathrm{F}}$ of a given configuration is thus obtained
as the maximum value of the external stress over the simulation:
\begin{equation}
\label{FlowStress}
\Sigma^{\mathrm{F}} = \max_t \Sigma^{\mathrm{ext}}(t)\;,
\end{equation}
where $t$ is an iteration step. For an external loading
$\Sigma^{\mathrm{ext}} < \Sigma^{\mathrm{F}}$, plastic deformation
will eventually stop while any loading $\Sigma^{\mathrm{ext}} <
\Sigma^{\mathrm{F}}$ will allow it to develop indefinitely.

\subsection{Application to amorphous composites}

The model presented above can be easily applied to the case of
amorphous composites. A major hypothesis (already performed in the
bare model) consists in assuming the homogeneity of the elastic
properties. Only the effect of a plastic disorder will be considered
in the following. To represent the amorphous composite we consider a
fraction $\phi$ of inclusions randomly distributed in an amorphous
matrix. Here the size of the inclusions is assumed to be given by the
mesh size so that no correlation is considered in the spatial
distribution of inclusions. The fraction of inclusions is defined by
$\phi=N_{\mathrm{inc}}/N^2$ where $N_{\mathrm{inc}}$ is the number
of hard  inclusions and $N$ the linear size of the lattice.

A bimodal distribution is used to account for the respective plastic
thresholds of the matrix and the inclusions.  For the amorphous
matrix, the plastic threshold is drawn from a uniform distribution
$[\overline{\sigma^{\mathrm{c}}}-\delta\sigma^{\mathrm{c}},
  \overline{\sigma^{\mathrm{c}}}+\delta\sigma^{\mathrm{c}}]$. Here we
choose $\overline{\sigma^{\mathrm{c}}}=1$ and
$\delta\sigma^{\mathrm{c}}=0.5$. The inclusions can be either less or
more ductile than the amorphous matrix. In the cases of interest
presented above, their nature is often crystalline. We thus assume low
fluctuations of the plastic properties of the inclusions and we
consider that they are characterized by a constant plastic threshold,
$\sigma^{\mathrm{c}}=\Sigma^{\mathrm{H}}$: all inclusions get the same
yield stress and this value does not change after an inclusion has
experienced plastic deformation. Here we restrict the scope to the case of hard
particles: $\Sigma^H > \overline{\sigma^{\mathrm{c}}}$. In order to
reduce the space of parameters we also consider that the typical
plastic strain undergone by the inclusions is the same as in the
amorphous matrix.

\subsection{Overview of the simulations}

Simulations were performed with sizes ranging from $N=16$ up to
$N=256$, and a number $M=40$ of independent realizations of the
disorder. The fraction of inclusions was varied between, $\phi=2.5 \times
10^{-4}$ and $\phi=0.99$. Different values of the contrast between
inclusions and matrix were used : $\Sigma^{\mathrm{H}}=4, 10, 40$ and the value
$\Sigma^{\mathrm{H}}=10^{8}$ was used to mimic infinitely hard particles.
Most of the following discussion will focus on the case $\Sigma^H = 10$.

\section{A size dependent effective yield stress\label{Size-effects}}

\subsection{Amorphous matrix}

We first discuss size effects in the case of a mere amorphous matrix,
i.e., in the absence of hard  particles. The ultimate yield
strength or flow stress $\Sigma^{\mathrm{F}}$ of the material is defined as the maximum stress
experienced by the material for a given simulation.

\begin{figure}
\begin{center}
  \includegraphics[width=0.98\columnwidth]{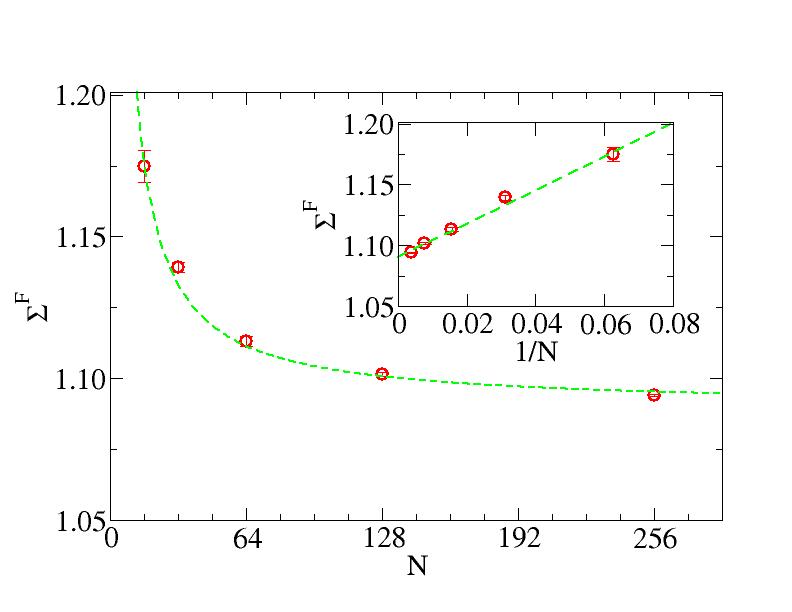}
\end{center} 
 \caption{\label{fig:size-effect-matrix} (color online) Variation of the
   ultimate yield strength $\Sigma^{\mathrm{F}}$ with the system size $N$ for a
   mere amorphous matrix ($\phi=0$) with a yield stress $\sigma^{\mathrm{c}}
   \in [0.5; 1.5]$. The line corresponds to the power
   law expression $\Sigma^{\mathrm{F}} = \Sigma^* + \frac{A}{N}$. As shown in the
   inset this evolution is consistent with the numerical data.}
\end{figure}

\begin{figure}
\begin{center}
  \includegraphics[width=0.98\columnwidth]{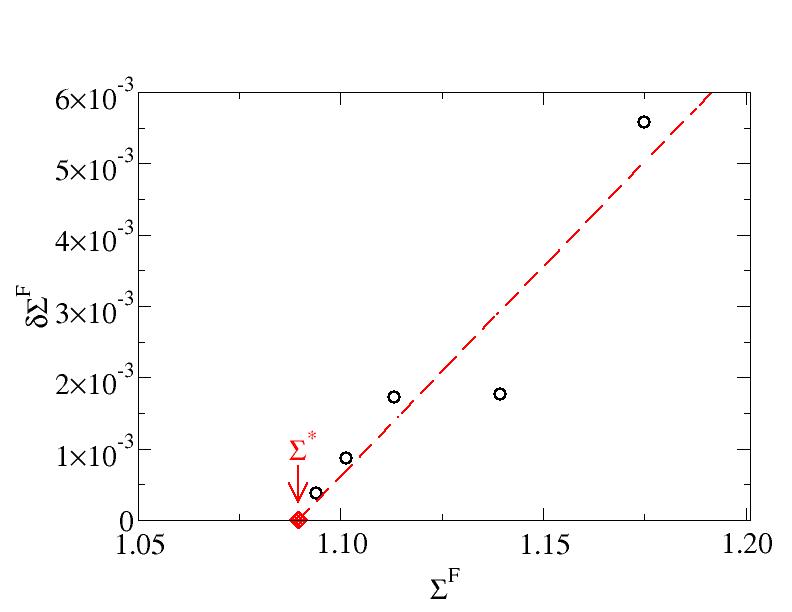}
\end{center} 
 \caption{\label{fig:sigmaVsNc0} (color online) Variation of the
   standard deviation $\delta\Sigma^{\mathrm{F}}$ of the 
   ultimate yield strength $\Sigma^{\mathrm{F}}$ with $\Sigma^{\mathrm{F}}$
   for the amorphous matrix with a yield stress $\sigma^{\mathrm{c}}
   \in [0.5; 1.5]$,
   for system sizes
   $N=16,\;32,\;64,\;128,\;256$. The standard deviation is obtained
   for $40$ realizations. As expected for a critical transition, a
   linear behavior is obtained. An extrapolation at zero standard deviation
   gives an estimate of the critical threshold, here the yield stress
   $\Sigma^*$ at infinite size.}
\end{figure}

In Fig.~\ref{fig:size-effect-matrix} we show the evolution of the
ultimate yield strength with the system size. A slight decrease is
observed. In the inset, we show that the evolution is consistent with a
simple power law dependence:
\begin{equation}
\Sigma^{\mathrm{F}} = \Sigma^* + \frac{A}{N}\;,
\end{equation}
where $\Sigma^*$ is the flow stress in the limit of an infinitely large system
and $A$ is a constant.
Such a power-law dependence is consistent with the depinning-like
nature of the model. In this
context~\cite{BVR-PRL02,TPVR-PRE11,Wyart-PNAS14}, the plastic flow
stress can be viewed as a critical threshold between a static phase
(no plasticity) and a dynamic phase (plastic flow). The fluctuations
of the depinning threshold measured over a finite length scale here
simply reflect the divergence of the correlation length in the
vicinity of a critical threshold $\xi \approx |f-f^*|^{-\nu}$. The
present results are consistent with the rough estimate $\nu\approx 1$
obtained in previous
works~\cite{TPVR-PRE11,Wyart-PNAS14}. Fig.~\ref{fig:sigmaVsNc0} gives
another illustration of this critical-like behavior. This figure shows
the variation of the standard deviation $\delta \Sigma^{\mathrm{F}}$ with the average
flow stress $\Sigma^{\mathrm{F}}$. The variation is reasonably
reproduced by an affine relationship $\delta \Sigma^{\mathrm{F}} =
a(\Sigma^{\mathrm{F}} - \Sigma^*)$. This is consistent with the
expected critical behavior $(\Sigma^* - \Sigma^{\mathrm{F}} ) \propto
\delta \Sigma^{\mathrm{F}} \propto L^{-1/\nu}$. The intercept value
$\Sigma^*$ can be seen here as the extrapolated value of the effective
flow stress at infinite size.

Note that, independently of the system size, the values of the
effective flow stress lie significantly above the simple average of
the microscopic thresholds $\overline{\sigma^{\mathrm{c}}}=1$.

\subsection{Amorphous composites}

Second, we discuss the dependence of the ultimate yield strength on the
fraction of inclusions and on the size of the system.

\begin{figure}
\begin{center}
  \includegraphics[width=0.98\columnwidth]{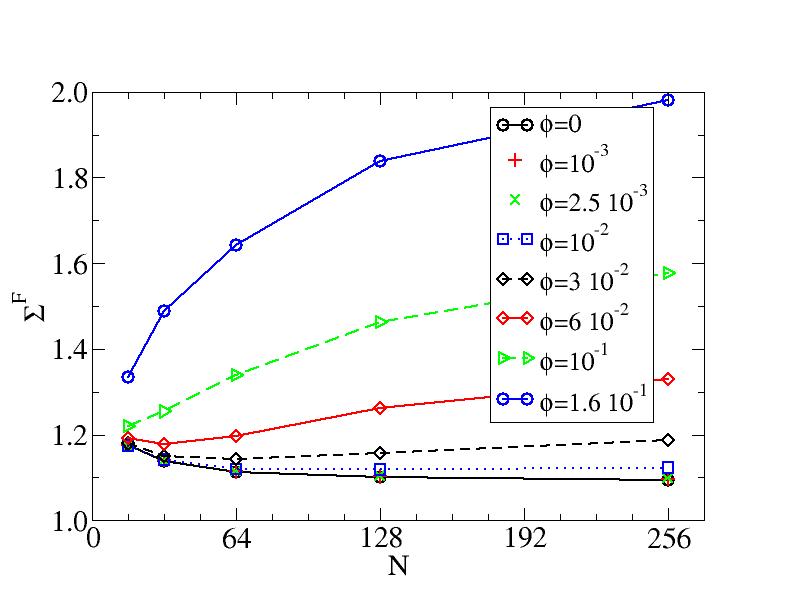}
\end{center} 
 \caption{\label{fig:sigmaVsN} (color online) Variation of the
   ultimate yield strength $\Sigma^{\mathrm{F}}$ with the system size $N$ for the
   amorphous matrix ($\phi=0$) and amorphous composites with various
   fractions of hard inclusions. 
   The yield stress of the amorphous matrix is $\sigma^{\mathrm{c}}
   \in [0.5; 1.5]$, the yield stress of the inclusions is $\Sigma^{\mathrm{H}} = 10$.
   Depending on $\phi$ the yield strength
   shows either a decreasing or an increasing trend with increasing system
   size.}
\end{figure}

{\em Size dependence \---} In Fig.~\ref{fig:sigmaVsN}, we show the size dependence observed for
amorphous composites with volume fractions of inclusions ranging from
$\phi=0$ to $\phi=0.16$.

For low fractions of hard inclusions, the behavior is similar to that
obtained for the amorphous matrix. The yield strength decreases
with increasing system size and converges towards a finite value for large system
sizes. 

Surprisingly, the behavior is markedly different for large fractions of
inclusions: the ultimate yield strength increases with increasing system size. At
intermediate values of the fraction of inclusions, the evolution of the yield strength even appears
to be non-monotonic.

\begin{figure}
\begin{center}
  \includegraphics[width=0.98\columnwidth]{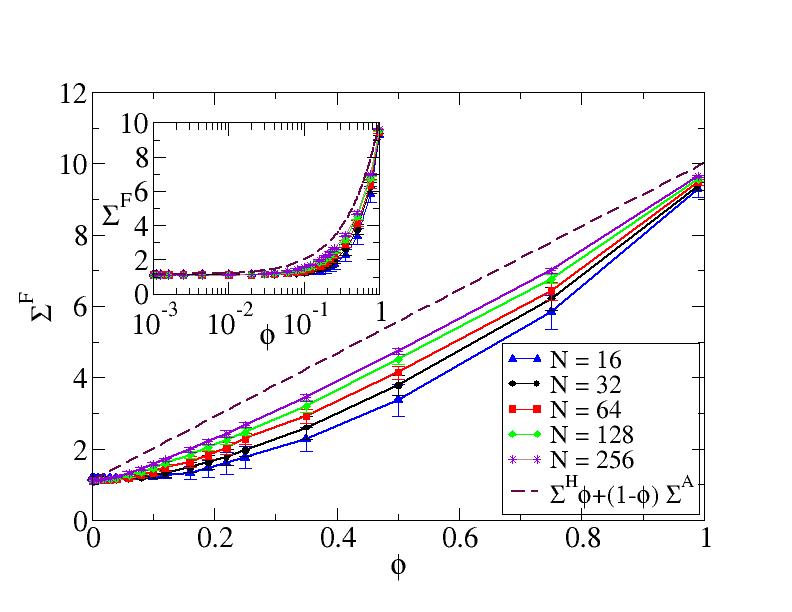}
\end{center} 
 \caption{\label{fig:sigmaVsConc} (color online) Variation of the
   ultimate yield strength $\Sigma^{\mathrm{F}}$ with the fraction
   $\phi$ of hard inclusions for a yield stress $\sigma^{\mathrm{c}}
   \in [0.5; 1.5]$ of the matrix, a yield stress $\Sigma^{\mathrm{H}}
   = 10$ of the hard inclusions and for five different system sizes
   $N=16$, $N=32$, $N=64$, $N=128$, and $N = 256$.  The same data are
   shown in the inset in semi-logarithmic scale.}
\end{figure}

{\em Mixing law \---}In Fig.~\ref{fig:sigmaVsConc} we show the
evolution of $\Sigma^{\mathrm{F}}$ with the fraction $\phi$ of inclusions of
yield stress $\Sigma^{\mathrm{H}}=10$ for system sizes ranging from $N=16$ to
$N=256$. The error bars indicate the standard deviation computed on
the different realizations performed for a given pair of parameters
$(\phi,N)$.

A clear size-effect is observed. The curves obtained for different
values of $N$ do not superimpose. The larger the system, the larger
the reinforcement effect induced by the hard inclusions and the closer
the effective yield strength to the value obtained from a simple
linear mixing law:
\begin{equation}
\label{MixingLaw}
\Sigma^{\mathrm{M}}(\phi, N) = (1-\phi)\Sigma^{\mathrm{A}}(N) +
\phi \Sigma^{\mathrm{H}}\;,
\end{equation}
where $\Sigma^{\mathrm{A}}$ is the ultimate yield strength of the sole
amorphous matrix and $\Sigma^{\mathrm{H}}$ the yield stress of the
hard sites.  Note that the value $\Sigma^{\mathrm{M}}$ obtained from a linear mixing law,
known as the Voigt average in
the context of homogenization is usually expected to be an upper
bound~\cite{Torquato-book02}. While this statement is true for
homogenization of linear properties such as conductivity or
elasticity, it does not necessarily hold for non-linear properties such
as fracture or plasticity. In the latter case, out-of-equilibrium
mechanisms may allow the effective property to reach values above the
Voigt bound~\cite{RVH-EJMA03,PVR-PRL13}.

Although it often fails to reproduce quantitatively the
experimental data, the simple linear mixing law~\cite{Leidner-JAPS74} remains
widely used in material science to account for the effects of plastic
reinforcement~\cite{Turcsanyi-JMSL88,Chen-JMCE98,Chen-JMS05}.

Another feature, here emphasized in the inset of
Fig.~\ref{fig:sigmaVsConc}, can be pointed out: for a given system size
$N$, no reinforcement is observed below a threshold value $\phi_c(N)$ of
the volume fraction of hard inclusions. The larger the system size
$N$, the smaller the threshold value $\phi_c(N)$.

Despite its simplicity (scalar model, perfect plasticity), the present
model is thus characterized by a complex behavior. In particular it
exhibits a clear size effect that can usually only be reproduced in
the framework of more complex descriptions of plasticity such as
strain-gradient based theories~\cite{Willis-JMPS04}. A key ingredient
is here the account of the elastic interaction induced by the local
plastic events.

\section{Hardening and localization\label{Hardening}}

We now discuss in more details the plastic behavior of the model
amorphous composites. In the following, we try to unveil the mechanisms
at play in the hardening regime. We shall discriminate between two
different effects, respectively associated to a structural evolution of
the amorphous matrix and a concentration of the stresses on the hard particles. We
then show a gradual localization of the plastic
deformation on the weakest band of the material.

\subsection{Stress-Strain curves}

\begin{figure}
   {\includegraphics[scale=0.25]{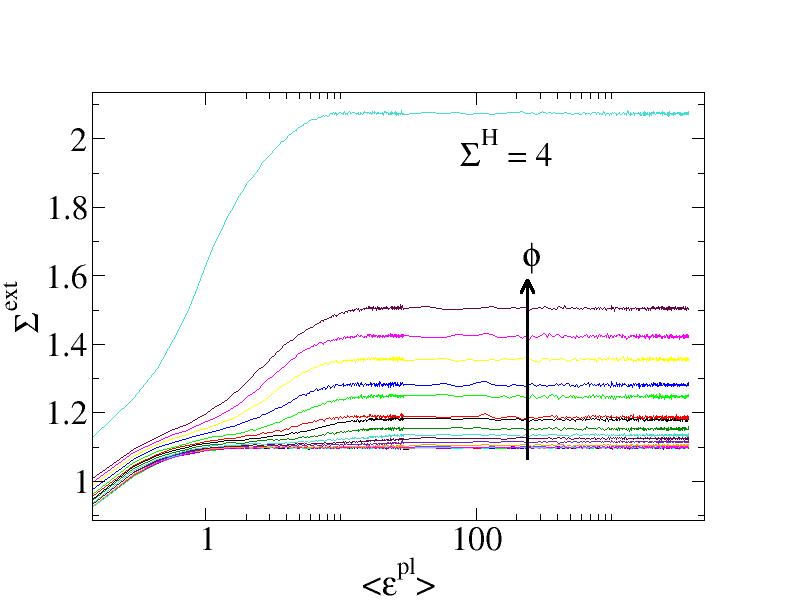}}
   \vspace{-4pt}
   {\includegraphics[scale=0.25]{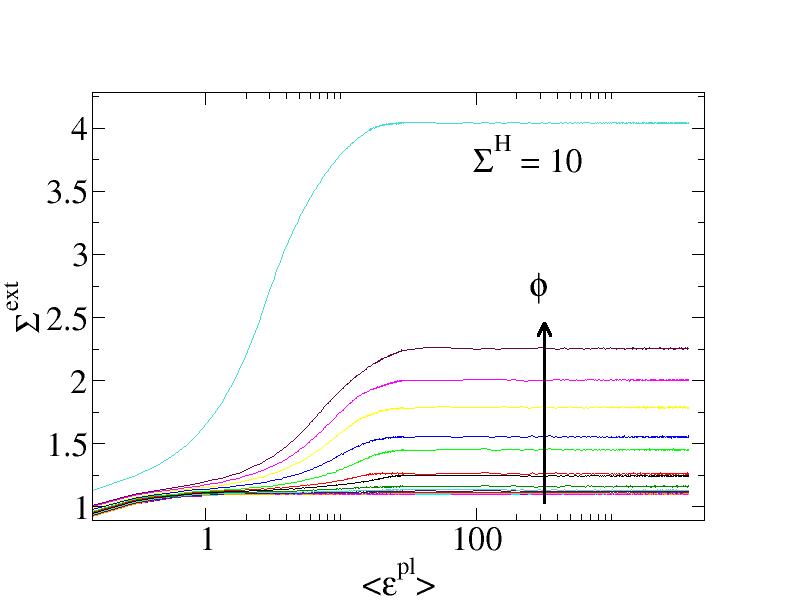}} 
   \vspace{-4pt}
   {\includegraphics[scale=0.25]{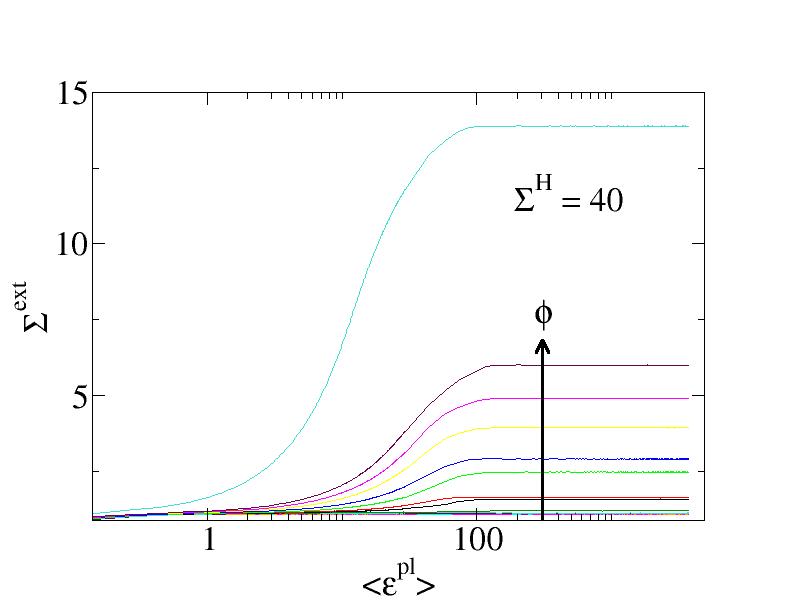}}  
   \vspace{-4pt}
   {\includegraphics[scale=0.25]{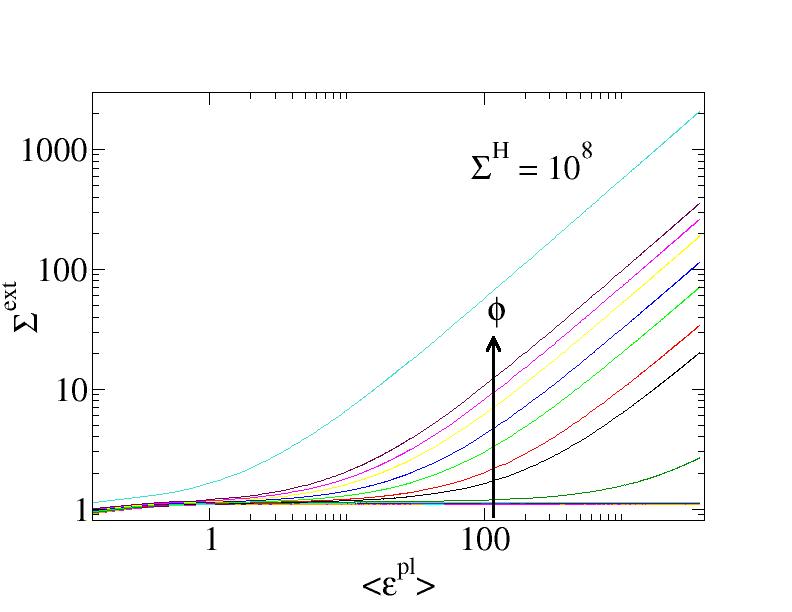}}
  \caption{\label{fig:stressStrain}
(color online) Stress-strain curves for a yield stress of the matrix $\sigma^{\mathrm{c}} \in [0.5; 1.5]$,
volume fractions of hard sites $\phi = \{ 0, ... ,0.25\}$ and for
a yield stress of the hard inclusions of $\Sigma^{\mathrm{H}} = 4$ (a),
$\Sigma^{\mathrm{H}} = 10$ (b), $\Sigma^{\mathrm{H}} = 40$ (c), and $\Sigma^{\mathrm{H}} = 10^8$ (d).}
\end{figure}

Figure~\ref{fig:stressStrain} displays the stress-strain curves
obtained for four different values of the inclusion yield stress
$\Sigma^{\mathrm{H}}=4,\;10,\;40,\;10^8$ (the latter case being meant to mimic
infinitely hard inclusions) and for different volume fractions $\phi$
ranging from $0$ to $0.25$.  Note that in order to emphasize the
hardening regime the variation of the stress is represented versus the
sole plastic strain.

Two successive hardening regimes can be distinguished before the
stress plateau corresponding to the flow stress is reached. The first
one is related to the hardening of the amorphous matrix. The second
one is directly induced by the presence of hard particles.

\subsection{Statistical hardening of the amorphous matrix}

\begin{figure}
\begin{center}
  \includegraphics[width=0.98\columnwidth]{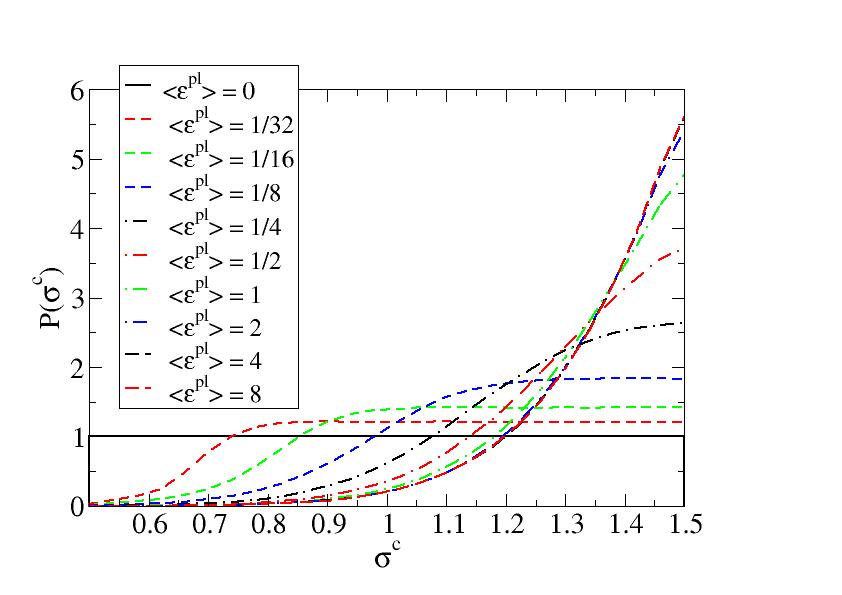}
\end{center} 
 \caption{\label{fig:trap-matrix-exhaustion} (color online) Evolution of the
   distribution of local plastic thresholds $P(\sigma^{\mathrm{c}})$ upon plastic
   deformation. A gradual exhaustion effect is observed until a
   stationary distribution is reached.}
\end{figure}

In this subsection, hardening in the pure amorphous matrix is
considered. At low plastic strain, a gradual hardening of the
amorphous matrix takes place. This phenomenon which has been
discussed in Refs.~\cite{BVR-PRL02,TPVR-CRM12} results from the
progressive exhaustion of the weakest sites of the matrix. We show in
Fig.~\ref{fig:trap-matrix-exhaustion} the gradual evolution of the
distribution of the local plastic thresholds $P(\sigma^{\mathrm{c}})$ upon
deformation in the case of the sole matrix. The larger the
deformation, the narrower the distribution and the closer the mean to
the upper bound value.  

This structural evolution can be understood in the following way. After plastic
rearrangements, the sites are given a new plastic threshold drawn from
the same random distribution as the initial ones. The systematic bias
between the weak thresholds of the failing sites and the ``normal''
thresholds that replace them after deformation induces an
evolutionary-like transient increase, reminiscent of self-organized
criticality models~\cite{BakSneppen-PRL93}. It can be seen 
in Fig.~\ref{fig:stressStrain} that at low
fractions of inclusions, this exhaustion mechanism is the only one to
hold and hard particles do not contribute to the reinforcement.
Indeed the stress-strain curve at low volume fractions of
hard sites is identical to that of the pure amorphous matrix.

\begin{figure}
\begin{center}
  \includegraphics[width=0.98\columnwidth]{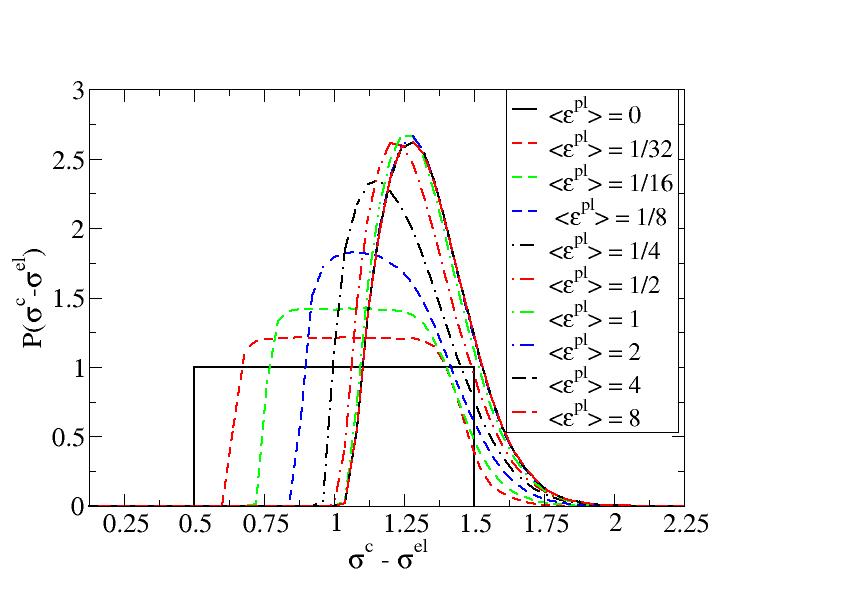}
\end{center} 
 \caption{\label{fig:crit-matrix-exhaustion} (color online) Evolution
   of the distribution of {\it effective} plastic thresholds
   $P(\sigma_c^{\mathrm{eff}})$ where $\sigma_c^{\mathrm{eff}} =\sigma^{\mathrm{c}} -
   \sigma^{\mathrm{el}} $
   upon plastic deformation. The sharp lower front is associated to
   the emergence of a global yield stress. The latter gradually
   increases upon plastic deformation (hardening) until it reaches its
   final value (the stress plateau of the stress-strain curve).}
\end{figure}

A complementary view of this statistical hardening is given in
Fig.~\ref{fig:crit-matrix-exhaustion}. Here, instead of the local
plastic thresholds, we show the evolution of the distribution of the
effective thresholds $P(\sigma_c^{\mathrm{eff}})$ where
$\sigma_c^{\mathrm{eff}} =\sigma^{\mathrm{c}} -
\sigma^{\mathrm{el}}$. Indeed, following Eq.~\ref{criterion}, the
local criterion for a given site $(i,j)$ can be rewritten as
$\Sigma^{\mathrm{ext}} \leq \sigma_{ij}^{\mathrm{c}} -
\sigma_{ij}^{\mathrm{el}}$. In other terms, the local thresholds are
dressed by the internal stress. Following the evolution of the
distribution upon deformation, we recover the hardening
effect. Interestingly, even in the transient stage, one can
identify a sharp front associated to the lower bound of the
distribution. This directly corresponds to the emergence of a yield
stress. The disordered system has self-organized and in the transient
stage one can unambiguously define a yield stress that depends on an
internal variable, the cumulated plastic strain. This also shows the
dependence of the macroscopic plastic properties on the past
mechanical history.

\subsection{Inclusion hardening}

When hard particles are present in the amorphous matrix, an additional
hardening stage is observed. As seen in in
Fig.~\ref{fig:stressStrain}, this second stage takes place at higher
plastic strains than the matrix hardening stage. We observe that the
higher the fraction of hard particles, the sooner the onset of the
second hardening regime. Note also that below a certain fraction of
hard inclusions, no second stage of hardening is observed.

In comparison to the pure amorphous matrix, the initial distribution
of plastic thresholds is bimodal in a composite. In the initial stage of the deformation, due to the
high contrast of plastic thresholds, only sites of the amorphous matrix
can experience plasticity. The plastic events induce internal
stresses. Hard particles can sustain a level of internal stress much
higher than that of the amorphous matrix and act here as a kind of internal
skeleton bearing most of the stress exerted on the structure.

\begin{figure}
\begin{center}
  \includegraphics[width=0.98\columnwidth]{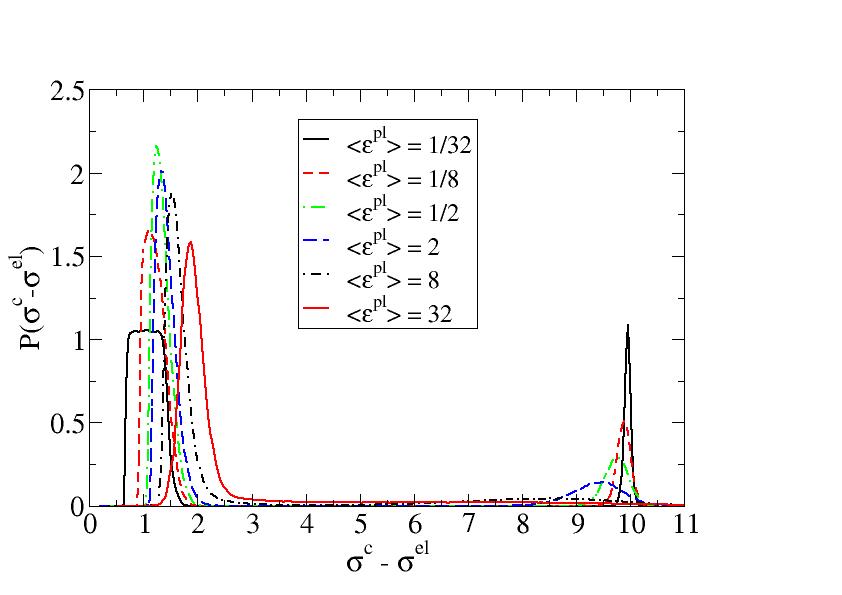}
\end{center} 
 \caption{\label{fig:crit-exhaustion-composite16} (color online) Evolution
   of the distribution of {\it effective} plastic thresholds
   $P(\sigma_c^{\mathrm{eff}})$ where $\sigma_c^{\mathrm{eff}} =\sigma^{\mathrm{c}} -
   \sigma^{\mathrm{el}} $
   upon plastic deformation in an amorphous composite with 16\% of hard
   particles of yield stress $\Sigma^{\mathrm{H}}=10$. The building of internal
   stresses gradually smears out the peak associated with hard
   particles. Conversely, the sharp front corresponding to the global
   yield stress increases upon plastic deformation longer than in the
   case of the sole amorphous matrix.}
\end{figure}

Again it is of interest to follow the distribution of effective
thresholds. In Fig.~\ref{fig:crit-exhaustion-composite16} we show the
evolution observed for an amorphous composite with 16\% of hard particles
of yield stress $\Sigma^{\mathrm{H}}=10$. We see that upon plastic deformation,
the build-up of internal stress on hard particles has a clear effect:
it tends to smear out the peak around $\Sigma^{\mathrm{H}}=10$. In the mean time
the lower part of the distribution, in particular the sharp front that
corresponds to the global yield stress keeps on increasing. This
second hardening stage is much longer than the statistical hardening
of the amorphous matrix. Stationarity is eventually obtained when the
second peak has entirely disappeared.

\subsection{Localization: no-slip bands}
\label{sec:localization}
In order to reveal the hardening mechanisms induced by the hard particles,
we give a closer look at the spatial organization of the plastic
strain field. In Fig.~\ref{fig:trap_rel_strain_maps} we show in the
top row maps of the relative plastic strain
$\varepsilon^{\mathrm{pl}}_{i, j}/\langle\varepsilon^{\mathrm{pl}}\rangle$
obtained after a long simulation for three concentrations of particles ($\phi=10^{-3}$,
$10^{-2}$, $10^{-1}$) and in the bottom row the associated maps of
plastic thresholds $\sigma^{\mathrm{c}}$ (in the final configuration
of a long simulation) indicating the position of the hard sites.

The low concentration case (panels (a) and (d) of
Fig.~\ref{fig:trap_rel_strain_maps}, $\phi=10^{-3}$) gives a good
illustration of the effect of adding hard sites on plastic deformation. We see
that the plastic strain field is not homogeneous. In this example
where only $3$ hard particles are present, we observe, as expected, that
the hard particles are barely deformed.
Interestingly, plastic deformation is also small along the
bands at $\pm 45^\circ$ that intercept the hard sites. Plasticity is
inhibited along a set of ``no-slip'' bands induced by the presence of
hard particles.  These bands orientated at $\pm45^\circ$ obviously
reflect the symmetry of the quadrupolar elastic interaction discussed
above. While the low fraction of hard inclusions shown in this example
is not sufficient to induce any reinforcement, it gives a simple clue
on the strengthening mechanism: hard particles inhibit the natural slip
systems associated to the elastic kernel~\cite{TPRV-preprint15}.

In the medium concentration case (panels (b) and (e), $\phi=10^{-2}$),
the (relative) plastic strain field is more heterogeneous. One
recovers patterns orientated at $\pm 45^\circ$ and it is possible to
distinguish between two kinds of bands: bands containing hard sites
are much less deformed than those not containing hard sites. In other
words, the lattice of no-slip bands is much denser and only the sites
not intercepted by these bands can easily undergo plastic deformation.

In the high concentration case (panels (c) and (f), $\phi=10^{-1}$),
the (relative) plastic strain field is highly heterogeneous and
actually highly localized. Most of the plastic deformation
concentrates onto one single band. This evolution is more clearly
shown in Fig.~\ref{fig:strain_maps_Phi01} where we represented maps of
the incremental plastic strain $\Delta \varepsilon^{\mathrm{pl}}_{i,j}
= \varepsilon^{\mathrm{pl}}_{i,j}(t + \delta t) - \varepsilon^{\mathrm{pl}}_{i,j}(t) $
where $\delta t$ represents a few iteration steps such that
$\langle \varepsilon^{\mathrm{pl}}\rangle (\delta t) = 2$
and $\langle \varepsilon^{\mathrm{pl}}\rangle (t) = 10$, $20$, $30$, $40$, $50$ and $60$.
Upon deformation, plastic activity
appears to become more and more localized.

\begin{widetext}

\begin{figure}[t]
\begin{center}
\includegraphics[height=0.3\textwidth]{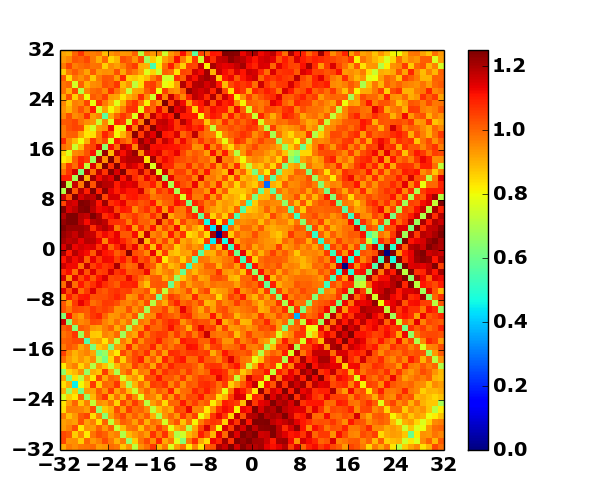}\hspace{-0.75cm}
\includegraphics[height=0.3\textwidth]{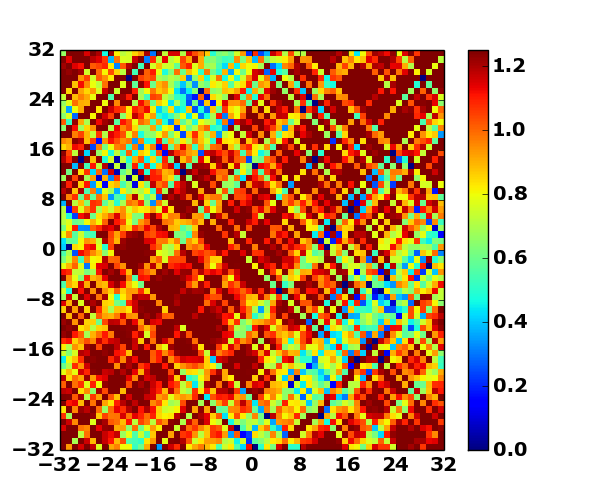}\hspace{-0.75cm}
\includegraphics[height=0.3\textwidth]{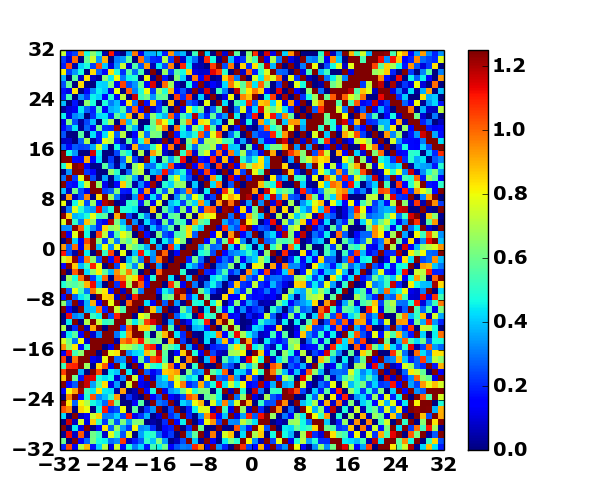}
\end{center}
\vspace{-0.5cm}
\hspace{-0.9cm}(a)\hspace{4.375cm}(b)\hspace{4.375cm}(c)\\
\vspace{-0.25cm}
\begin{center}
\includegraphics[height=0.3\textwidth]{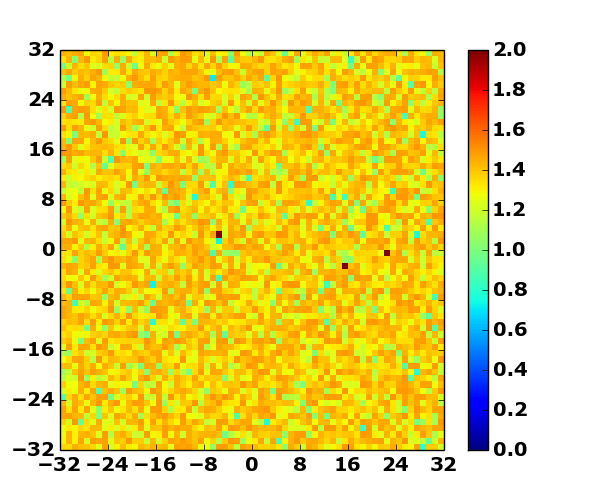}\hspace{-0.75cm}
\includegraphics[height=0.3\textwidth]{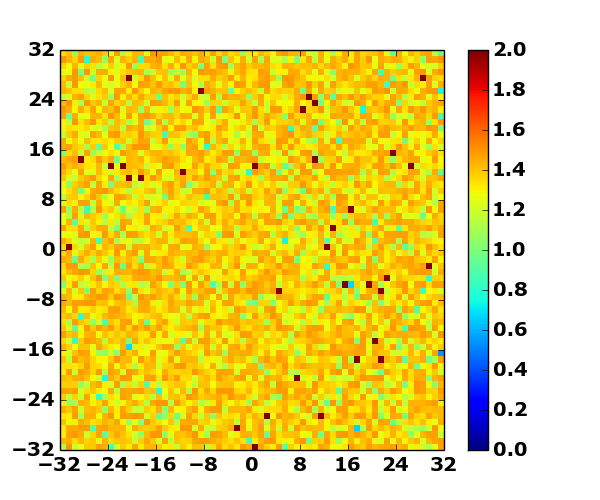}\hspace{-0.75cm}
\includegraphics[height=0.3\textwidth]{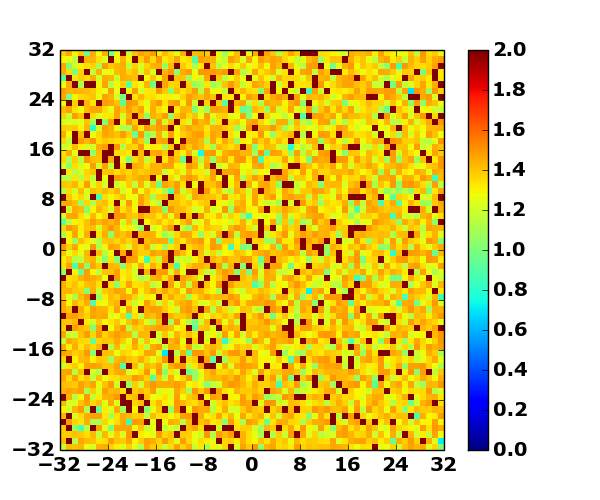}
\end{center}
\vspace{-0.5cm}
\hspace{-0.9cm}(d)\hspace{4.375cm}(e)\hspace{4.375cm}(f)\\
\caption{\label{fig:trap_rel_strain_maps} (color online) (a),(b),(c):
  maps of the relative plastic strain $\varepsilon^{\mathrm{pl}}_{i, j}/\langle
  \varepsilon^{\mathrm{pl}}\rangle$ for 
   a system size $N=64$, a yield stress of hard sites $\Sigma^{\mathrm{H}} = 10$, and
   volume fractions of hard inclusions $\phi
   = 10^{-3}$, $10^{-2}$, $10^{-1}$ of hard sites,respectively.
  (d),(e),(f): maps of the associated final configurations of plastic
  thresholds $\sigma^{\mathrm{c}}$ . The positions of hard sites are visible in dark red.  }
\end{figure}

\begin{figure}[t]
\begin{center}
\includegraphics[height=0.3\textwidth]{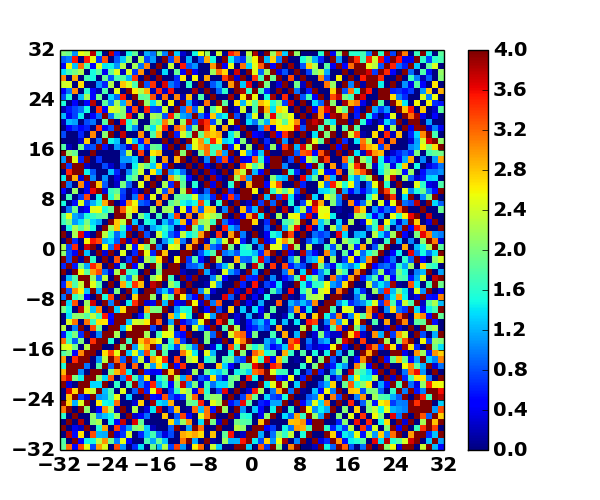}\hspace{-0.75cm}
\includegraphics[height=0.3\textwidth]{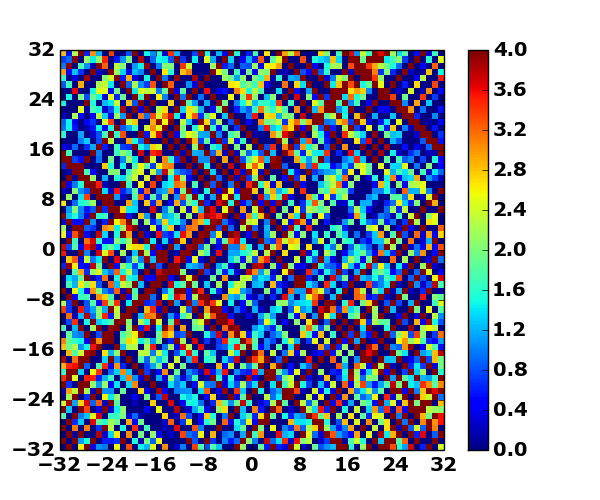}\hspace{-0.75cm}
\includegraphics[height=0.3\textwidth]{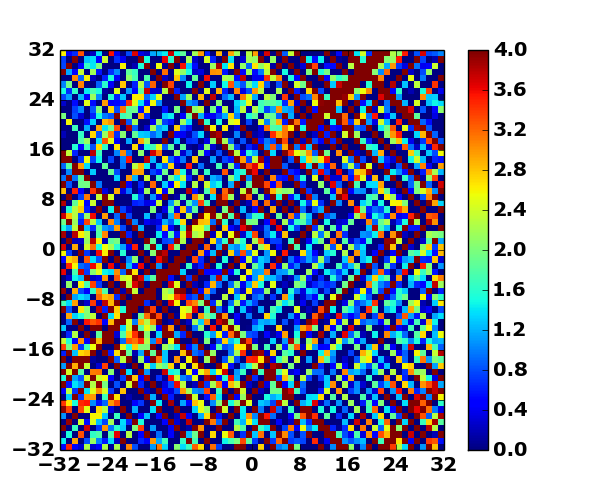}
\end{center}
\vspace{-0.5cm}
\hspace{-0.9cm}$\langle  \varepsilon^{pl}\rangle=10$
\hspace{3.375cm}$\langle  \varepsilon^{pl}\rangle=20$
\hspace{3.375cm}$\langle  \varepsilon^{pl}\rangle=30$\\
\vspace{-0.25cm}
\begin{center}
\includegraphics[height=0.3\textwidth]{fig12c.png}\hspace{-0.75cm}
\includegraphics[height=0.3\textwidth]{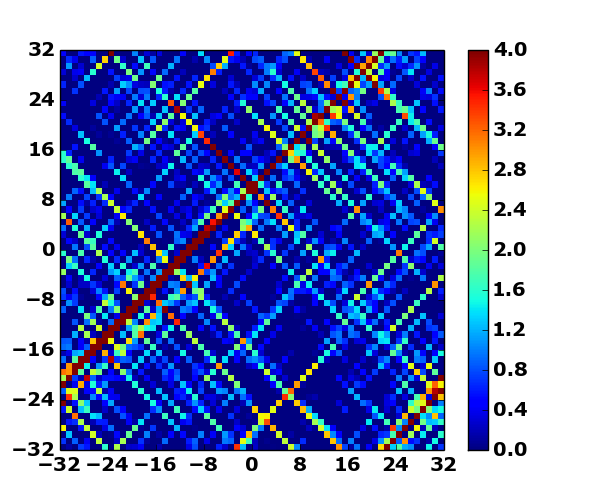}\hspace{-0.75cm}
\includegraphics[height=0.3\textwidth]{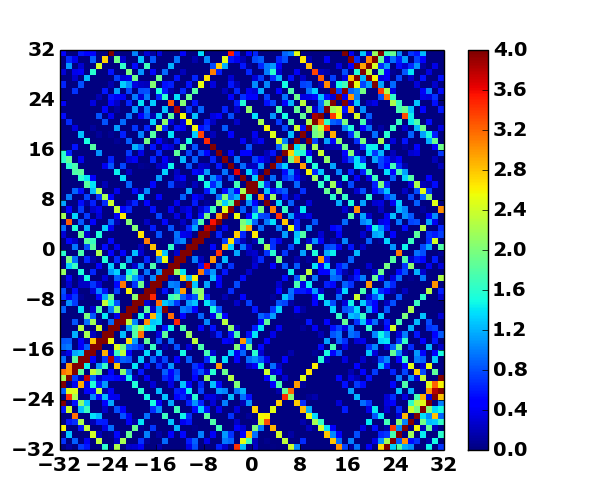}
\end{center}
\vspace{-0.5cm}
\hspace{-0.9cm}$\langle  \varepsilon^{pl}\rangle=40$
\hspace{3.375cm}$\langle  \varepsilon^{pl}\rangle=50$
\hspace{3.375cm}$\langle  \varepsilon^{pl}\rangle=60$\\
\caption{\label{fig:strain_maps_Phi01} (color online) maps of
  incremental plastic strain $\Delta\varepsilon^{\mathrm{pl}}$ for a volume
  fraction of hard inclusions $\phi = 10^{-1}$, a system size $N= 64$,
  a yield stress of hard sites $\Sigma^{\mathrm{H}} = 10$, and
  different values of the average plastic strain $\langle  \varepsilon^{pl}\rangle$.  }
\end{figure}

\end{widetext}

\subsection{Localization: the weakest band}
We now try to correlate the plastic activity with the underlying
structure, here represented by the landscape of plastic thresholds. As
discussed above, plastic deformation tends to localize along bands
orientated at $\pm 45^\circ$ that reflect the symmetry of the Eshelby
quadrupolar elastic interaction. Due to statistical fluctuations, not
all possible slip systems encounter the same number of hard
particles. We define the weakest band $SB_{\mathrm{min}}$ and the
strongest bands $SB_{\mathrm{max}}$ as the bands containing respectively the
smallest and the largest amount of hard particles among the $2N$ possible
slip systems.  Here we take into account the two possible
orientations. Note again that we consider periodic boundary
conditions so that all slip systems are a priori \textit{equivalent}. We can now
compute the fraction of plastic activity occurring in the various
bands. In order to highlight the gradual development of the localization,
we proceed as in Sec.~\ref{sec:localization}: we consider the evolution
of the incremental plastic strain field 
 $\Delta \varepsilon^{\mathrm{pl}}_{i,j}
= \varepsilon^{\mathrm{pl}}_{i,j}(t + \delta t) - \varepsilon^{\mathrm{pl}}_{i,j}(t) $
with $\delta t$ a few iteration steps such that
$\langle \varepsilon^{\mathrm{pl}}\rangle (\delta t) = 2$.

\begin{figure}[t]
\begin{center}
\includegraphics[width=0.98\columnwidth]{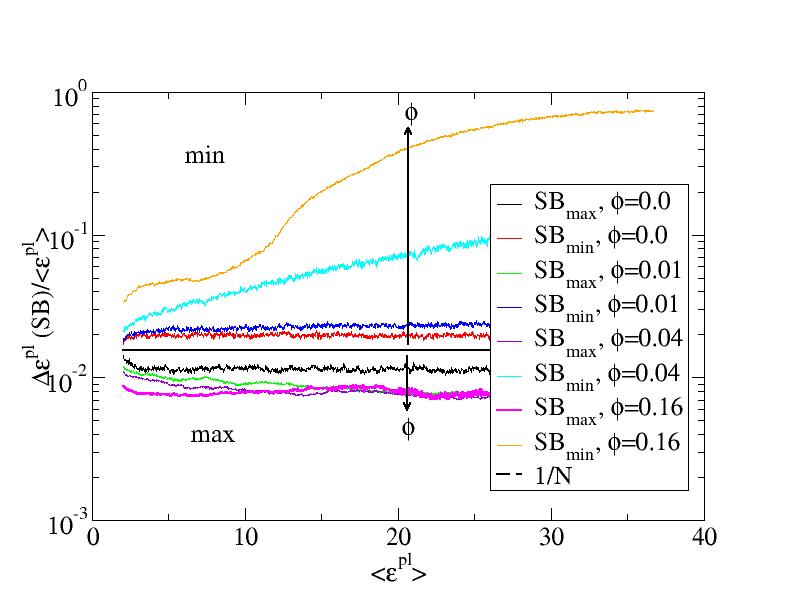}\hspace{-0.75cm}
\end{center}
\caption{\label{fig:fracSB10} Fractions $f_{\mathrm{min}}$ and $f_{\mathrm{max}}$ of incremental plastic strain $\Delta \varepsilon^{\mathrm{pl}}$  borne by the weakest and the strongest slip systems $SB_{\mathrm{min}}$ and $SB_{\mathrm{max}}$ containing the smallest and the largest amount of particles, respectively.}
\end{figure}

In Fig.~\ref{fig:fracSB10} we show the evolution with the cumulated
plastic strain $\langle \varepsilon^{\mathrm{pl}}\rangle$ of the fractions
$f_{\mathrm{min}} =  \Delta \varepsilon^{\mathrm{pl}} (SB_{\mathrm{min}})/ \langle \Delta \varepsilon^{\mathrm{pl}}\rangle$
and $f_{\mathrm{max}} = \Delta \varepsilon^{\mathrm{pl}} (SB_{\mathrm{max}})/ \langle \Delta \varepsilon^{\mathrm{pl}}\rangle$
of the incremental plastic strain borne by the weakest and the strongest bands
respectively for different concentrations of hard particles. If
the plastic strain field was uniformly spread on all bands, one would expect $f_{\mathrm{min}} =
f_{\mathrm{max}} =1/N$ (and not $1/2N$ because of the redundancy between the
two possible orientations at $+45^\circ$ and $-45^\circ$).

In the case of the sole amorphous matrix, the weakest band deforms about twice as more than the
strongest band: $f_{\mathrm{min}} \approx 2 f_{\mathrm{max}}$. For a fraction
$\phi=10^{-2}$, the effect is a bit more pronounced but not spectacular
yet, we observe $f_{\mathrm{min}} \approx 4 f_{\mathrm{max}}$. This ratio remains
reasonably constant upon deformation. This is consistent with the
typical heterogeneity observed in
Fig.~\ref{fig:trap_rel_strain_maps}. Note that for such a concentration
the number of particles falls strictly below the number of slip
systems so that deformation can always find a band free of particles
to develop. No significant reinforcement is expected in this case.

Above some threshold, all slip systems are virtually blocked by
hard particles. This is the case for the two concentrations $\phi=0.01$ and
$\phi=0.16$ shown in Fig.~\ref{fig:fracSB10}. Here we see a dramatic
effect: upon deformation, the weakest band bears a higher and higher
fraction of the plastic activity. Eventually most of the plastic
strain occurs within this weakest band. We thus observe a strong
correlation between structure and plastic behavior: plastic
deformation gradually concentrate onto the weakest slip system,
characterized by the smallest amount of hard particles.

\section{A simple analytical model\label{Analytical-Model}}

The mechanism of reinforcement can thus be understood in the following
way. Hard particles inhibit slip systems. No reinforcement occur until
all slip systems are blocked. Above the associated threshold
concentration, all slip systems are hindered by hard particles and
plastic strain gradually localizes onto the weakest one, i.e., the one that
contains the fewest hard particles. The macroscopic plastic behavior is thus
controlled by the properties of this weakest band. In the following, we
discuss these two aspects, elaborate a simple analytical model and
compare its prediction with our simulations. Mathematical details are
presented in a separated appendix.

\subsection{Percolation}
As discussed above, no reinforcement is expected until all slip systems
are blocked by at least one particle. Here the two families of slip
systems associated to the two directions at $\pm45^\circ$ should a
priori be considered. For the sake of simplicity, we consider in the
following only one of the two orientations. This approximation allows
to recover a simple one-dimensional percolation problem.

\begin{figure}
  \includegraphics[width=0.95\columnwidth]{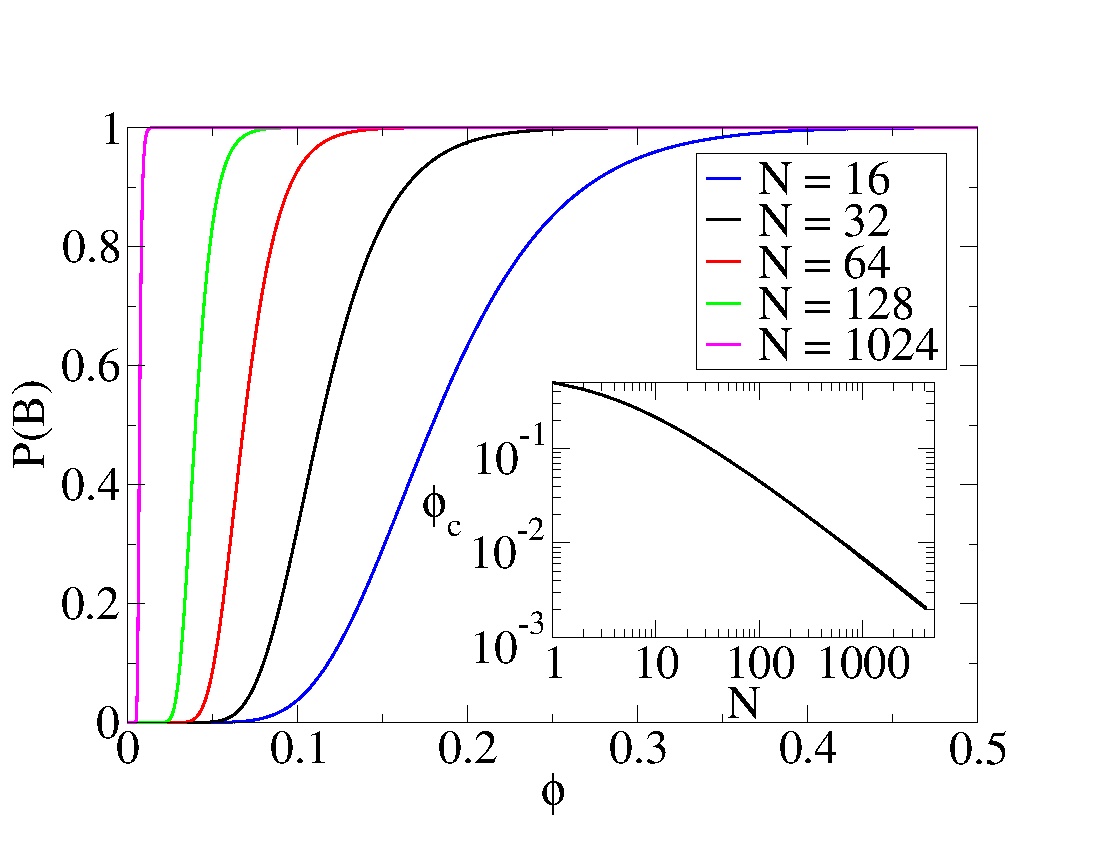}
  \caption{\label{fig:probaBlocked} (color online) Variation of the
    probability $P(B)$ of having at least one hard inclusion on each
    diagonal, defined in Eq.~\ref{eq:probabBlocked}, with the volume
    fraction $\phi$ of hard inclusions for different system sizes
    $N=16$, $N=32$, $N=64$, $N=128$, $N=1024$. Inset: Variation of the
    critical fraction $\phi_c$ defined in Eq.~\ref{eq:phic} with the
    system size $N$.  }
\end{figure}

We assume here that the distribution of particles is not spatially
correlated and take the volume fraction $\phi$ as the probability for
one inclusion to be hard.  The probability to have exactly $n$ hard
inclusions in one randomly chosen diagonal is then
\begin{equation}
P(N_d = n) = \binom{N}{n} \phi^n (1-\phi)^{N-n}\;,
\end{equation}
where $N_d$ is the random variable counting the number of hard sites
on a diagonal, $N_d = n$ is the event ``$n$ hard sites on a
diagonal", $N$ is the number of sites in a diagonal, which is exactly
the system size in the square lattice considered here. We recognize a
binomial distribution. The probability of having at least $1$ hard
inclusion on a diagonal is:
\begin{equation}
P(N_d \ge 1) = 1- (1-\phi)^{N}\;.
\end{equation}
There are $N$ diagonals with the same orientation. They are independent. Consequently,
the probability to have at least $1$ hard inclusion on each diagonal is
\begin{equation}
\label{eq:probabBlocked}
P(B) = \Bigl (1- (1-\phi)^{N}\Bigr )^N\;,
\end{equation}
where the letter $B$ stands for ``blocked''.  This probability is the
equivalent of the probability of percolation.
It is plotted for different system sizes versus the volume fraction of
hard inclusions in Fig.~\ref{fig:probaBlocked}. The probability of
having at least one hard inclusion per diagonal increases with the
volume fraction of hard inclusions until it reaches $1$.

\begin{figure}
  \includegraphics[width=0.95\columnwidth]{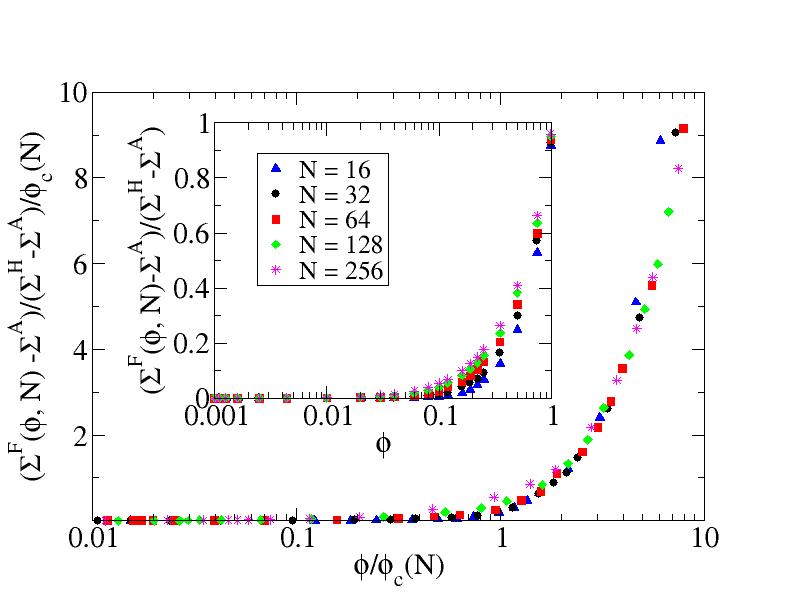}
  \caption{\label{fig:rescale} (color online) Variation of the
    rescaled yield strength $(\Sigma^{\mathrm{F}}(\phi,
    N)-\Sigma^{\mathrm{A}})/
    (\Sigma^{\mathrm{H}}-\Sigma^{\mathrm{A}})/\phi_c(N)$ with the
    rescaled volume fraction $\phi/\phi_c(N)$ for a yield stress
    $\Sigma^{\mathrm{H}} = 10$ of hard inclusions and different system
    sizes $N=16$, $32$, $64$, $128$, $256$.
    Inset: Variation of $(\Sigma^{\mathrm{F}}(\phi,
    N)-\Sigma^{\mathrm{A}})/
    (\Sigma^{\mathrm{H}}-\Sigma^{\mathrm{A}})$ with $\phi$ for the same systems.}
\end{figure}

The threshold for percolation, or here for all diagonals to be
blocked, is the volume fraction $\phi$ for which the probability
$P(B)$ is the steepest. In other words, the threshold for the
transition corresponds to the volume fraction of hard inclusions for
which the second derivative of $P(B)$ vanishes. This volume fraction
is called critical and denoted $\phi_c$.  It is equal to
\begin{equation}
\label{eq:phic}
 \phi_c(N) = 1-\frac{1}{(N+1)^{1/N}}\;.
\end{equation}
The inset of Fig.~\ref{fig:probaBlocked} shows the variation of the
critical fraction $\phi_c$ with the system size $N$. The thresholding
effect is nicely illustrated in Figure~\ref{fig:rescale} where we show
the rescaled ultimate yield strength
$(\Sigma^{\mathrm{F}}(\phi,
    N)-\Sigma^{\mathrm{A}})/
    (\Sigma^{\mathrm{H}}-\Sigma^{\mathrm{A}})/\phi_c(N)$ 
versus the rescaled volume fraction $\phi/\phi_c(N)$, for
different system sizes. This plot is to be compared with the inset
of Fig.~\ref{fig:rescale} showing the same quantities without
the rescaling by $\phi_c(N)$. In the main plot,
the curves corresponding to different system sizes collapse onto a
single master curve, showing that our interpretation of the transition
is valid.

\subsection{Yield stress of the weakest band}

\subsubsection{Restriction to elastic line depinning}
In section~\ref{Hardening}, plastic deformation was shown to
concentrate onto one single band, the one containing the smallest amount
of hard particles. It is thus natural to use the ultimate yield strength of that
weakest band as an estimate for the ultimate yield strength of the
whole amorphous composite. Ignoring the residual plastic strain
undergone outside the band, the problem is thus reduced to a
one-dimensional elastic depinning problem very similar to the
propagation of a crack front in a random
landscape~\cite{Schmittbuhl-PRL95,PVR-PRL13}.

Indeed, if we denote
$\varepsilon_i^{\mathrm{W}}=\varepsilon^{\mathrm{W}}(z_i)$ the plastic
strain in the weakest band at location $z_i$ where $z$ is the distance
along the band, any local plastic strain increment
$\delta\varepsilon_i^{\mathrm{W}}$ induces along the band an internal
stress proportional to an elastic kernel which is nothing but the
restriction on a diagonal of the Eshelby quadrupolar stress defined in
Eq.~\ref{quadrupolar-stress}. More specifically the internal stress at
location $z_j$ induced by the plastic strain increment at location
$z_i$ amounts to:
\begin{eqnarray}\nonumber
\delta\sigma_{ii}^{\mathrm{W}} &= & -A_0 \varepsilon_i^{\mathrm{W}}\;, \\
\delta\sigma_{ij}^{\mathrm{W}} &=&\frac{A_1}{(z_i-z_j)^2} \varepsilon_i^{\mathrm{W}}\;, \quad \mathrm{if} \quad i \ne j \;,
\end{eqnarray}
where $A_0$ and $A_1$ are positive constants. One recognizes here the
elastic interaction associated to the trapping of an interfacial crack
front~\cite{GaoRice-JAM89}.

The determination of the effective toughness of an interfacial crack
propagating in a random landscape (which also amounts to the critical
threshold of a long-range elastic line) has recently been discussed in
Ref~\cite{PVR-PRL13}. While the effective toughness can significantly
exceed the simple arithmetic average of the microscopic properties
when the disorder is highly fluctuating in the direction of
propagation (strong pinning), a simple mixing law is recovered when
the microscopic toughness is only slowly varying along the direction
of propagation (weak pinning).

In the present case, the hard sites are persistent, i.e., the value of
their yield stress does not change upon deformation. Besides, the
fluctuations of the local thresholds that characterize the amorphous
matrix are weak compared with the yield stress of the hard sites. Weak
pinning conditions can thus be considered and a simple mixing law used
to compute the effective yield stress of the band.

\subsubsection{How weak is the weakest band?}

The effective yield stress of the weakest band $ \sigma_Y^W$ is thus simply written:
\begin{equation}
\Sigma^{\mathrm{W}} = \frac{1-m}{N} \Sigma^{\mathrm{A}} + \frac{m}{N} \Sigma^{\mathrm{H}} \;,
\label{sigma_weakest_band}
\end{equation}
where $ \Sigma^{\mathrm{A}}$ is the yield stress of the amorphous matrix and $
\Sigma^{\mathrm{H}}$ that of the hard sites and $m$ is the number of hard sites
in the band.  The estimate of the ultimate yield strength $\Sigma^{\mathrm{F}}$ of the
material is given by the ensemble average:
\begin{equation}
\Sigma^{\mathrm{F}} = \langle \Sigma^{\mathrm{W}}  \rangle= \frac{1-\langle m\rangle}{N} \Sigma^{\mathrm{A}} + \frac{\langle m \rangle}{N} \Sigma^{\mathrm{H}} \;.
\label{sigma_weakest_band_avg}
\end{equation}
where $\langle m \rangle$ is the average minimum number of hard sites
on a diagonal of size $N$ for a fraction $\phi$ of hard sites. In the
following we define $f=\langle m \rangle/N$, the effective fraction of
hard sites in the weakest band. As it immediately appears from
Eq.~\ref{sigma_weakest_band_avg}, within the weakest band
approximation, the difference between the effective flow stress
$\Sigma^{\mathrm{F}}$ and the mixing law estimate
$\Sigma^{\mathrm{M}}$ directly stems from the difference between $f$
and $\phi$.

The distribution of the number $m$ of hard sites is given by the
binomial distribution of parameters $\phi$ and $N$.  An exact formula
for the average $\langle m\rangle$ of the minimal number of hard sites
on a diagonal when $N$ diagonals are considered is given in the
appendix. However, this formula contains an infinite sum and is not
easy to handle.  In order to estimate this minimal value we shall
resort to an argument of extremal statistics. Depending on the value
of $\phi$, the binomial converges at large $N$ either to a Gaussian or
to a Poisson distribution. In the present case we are interested in
the large deviations of the binomial
distribution~\cite{Arratia-BMB89}. We use recent results on the
general approximation of the binomial
distribution~\cite{Blondeau-PhD11,Blondeau-DCC11} obtained in the
context of cryptology studies:
\begin{equation}
\label{eq:blondeauf}
P(m \leq f N) =  \frac{\phi \sqrt{1-f}}{(\phi - f)\sqrt{2\pi N f}} e^{-N D(f||\phi)}\;,  
\end{equation}
for $N \rightarrow \infty$ where $D(f||\phi)$ is the Kullback-Leibler
divergence defined as:
\begin{equation}
D(f||\phi) = f \ln{\frac{f}{\phi}} + (1-f)\ln{\frac{1-f}{1-\phi}}\;.
\end{equation}
Here, the fraction $f$ of hard sites in the weakest band is estimated via a simple extremal statistics argument:
\begin{equation}
\label{eq:extremal-stat}
P(m \leq f N) \approx \frac{1}{N}\;.
\end{equation}
Detailed calculations based on the asymptotic expansions given in
Ref.~\cite{Blondeau-DCC11} are presented in the appendix. They allow
us to obtain an estimate of the distance between the fraction $f$ (the
fraction of hard sites in the weakest band) and the parameter $\phi$
of the binomial distribution (the mean fraction of hard sites):
\begin{equation}
\label{eq:resultf-main}
f = \phi - \sqrt{\frac{2\phi (1-\phi)}{N} \log \frac{N}{\sqrt{2\pi}}} (1+ r_N)\;,
\end{equation}
where
\begin{eqnarray}
\nonumber
r_N &= &- \frac{1}{2} \frac{\log (2h_N)}{2 h_N +1}\;,\\
h_N &= &\log \frac{N}{\sqrt{2\pi}} \;.
\end{eqnarray}

\begin{figure}
  \includegraphics[width=0.95\columnwidth]{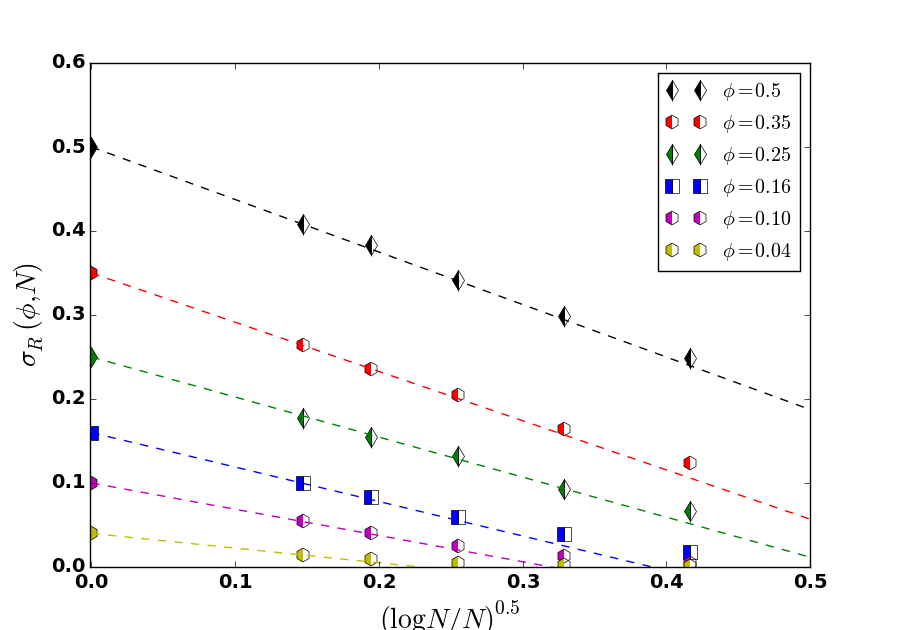}
  \caption{\label{size-scaling} (color online) Size scaling of the
    rescaled flow stress $\sigma_R(\phi,N)$, defined in Eq.~\ref{eq:rescaled-reinforcement},
    of amorphous composites with concentration of
    hard particles ranging from $\phi=0.04$ to $\phi=0.5$. Filled
    symbols on the vertical axis correspond to the infinite size
    limit, i.e., the result of the mixing law. Indicative dashed lines
    show the expected scaling behavior in $(\log N /N)^{1/2}$.}
\end{figure}

This immediately sets the distance of the flow stress
$\Sigma^{\mathrm{F}}$ to the mixing law value
$\Sigma^{\mathrm{M}}(\phi, N)$ obtained by Eq.~\ref{MixingLaw}:
\begin{equation}
\label{eq:result-vs-mixing-law}
\Sigma^{\mathrm{F}}(\phi, N) = \Sigma^{\mathrm{M}}(\phi, N) - \left(\phi - f \right) \left[\Sigma^{\mathrm{H}} - \Sigma^{\mathrm{A}}(N) \right]\;.
\end{equation}
In particular we obtain a clear size effect: the convergence to the
mixing law scales as $(\log N/N)^{1/2}$. This result is illustrated
in Fig.~\ref{size-scaling} where we display the variation of the
rescaled flow stress $\sigma_R(\phi,N)$ with $(\log N/N)^{1/2}$ for various values of
the fraction $\phi$ of hard sites.
The rescaled flow stress $\sigma_R(\phi,N)$ is defined as the reinforcement factor with respect to the flow stress of the amorphous matrix:
\begin{equation}
\label{eq:rescaled-reinforcement}
\sigma_R(\phi,N) = \frac{\Sigma^{\mathrm{F}}(\phi,N) -\Sigma^{\mathrm{A}}(N) }{\Sigma^{\mathrm{H}} - \Sigma^{\mathrm{A}}(N)}\;.
\end{equation}
In the framework of our approximation, we expect $\sigma_R(\phi,N) =
f((\phi,N)$. In particular, following Eq.~\ref{eq:resultf-main}, we
should recover $\phi - \sigma_R(\phi,N) \propto (\log N/N)^{1/2}$. As
shown in Fig.~\ref{size-scaling}, this scaling is nicely obeyed for
moderate values of $\phi$. Departures from the predicted scaling
behavior become significant at low values of $\phi$ and $N$,
because the analytical estimation holds only in the limit of large
$N$ and intermediate values of $\phi$. A numerical estimation of the
average number $\langle m \rangle$ of hard sites in the weakest band
is discussed in the appendix and shows that the approximation
holds surprisingly well even for low values of $\phi$ and small system
sizes.

\begin{figure}
  \includegraphics[width=0.95\columnwidth]{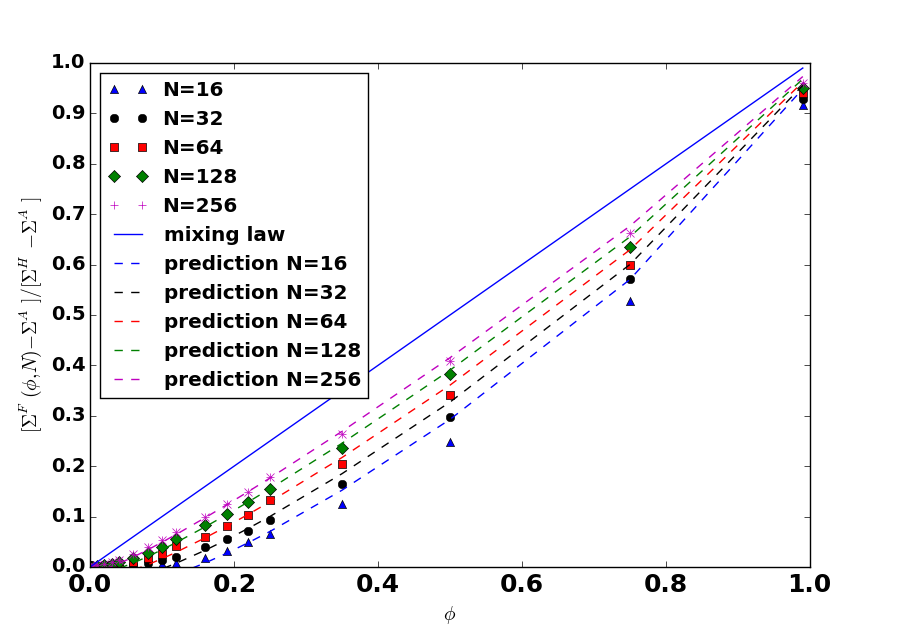}
  \caption{\label{ana-vs-num} (color online) Effect of the
    concentration of hard particles on the rescaled flow stress of
    amorphous composites for different system sizes $N=16$, $32$, $64$,
    $128$, $256$. The straight line corresponds to the mixing law
    expected at infinite size. The dashed lines are the analytical
    predictions of Eq.~\ref{eq:rescaled-reinforcement} accounting for
    logarithmic corrections.}
\end{figure}

Beyond the prediction of the scaling behavior, the logarithmic
corrections accounted for in Eq.~\ref{eq:resultf-main} allow us to
test quantitatively our predictions for the reinforcement effect of
hard inclusions in an amorphous matrix. In Fig.~\ref{ana-vs-num} we
compare analytical predictions and simulation results for the rescaled
flow stress $\sigma_R(\phi,N)$ with respect to the fraction of hard
sites $\phi$. Again, our prediction of effective flow stress happens
to be very precise for moderate values of $\phi$ and large system
sizes.

\section{Conclusion\label{Conclusion}}

The plastic behavior as described in the mesoscopic simulations shows
two types of system size dependence. The first type corresponds to an
effect of the amorphous matrix only and results from the critical
character of the yielding transition. In this case, the ultimate yield strength
decreases with an increasing system size, as $1/N$. 
This
system size dependence has already been addressed in Refs.~\cite{TPVR-PRE11,Wyart-PNAS14}.
A similar critical behavior has
recently been advocated in the related framework of compressive
strength of brittle heterogeneous materials~\cite{WGGAV-PNAS14}

The second type of size effect is specific to the composite
material. Below a critical volume fraction of hard inclusions
depending on the system size, no hardening behavior of the second type
is observed.  Above this critical volume fraction, the hardening
behavior depends on the system size: the ultimate yield strength increases
with an increasing system size, as $-(\log{N}/N)^{1/2}$.
We showed that the thresholding
effect observed in the simulations is close to a percolation
transition. We also showed that during this second hardening regime,
most of the plastic strain is concentrated onto the weakest band.
Therefore, we proposed a simple model to describe the dependence of
the ultimate yield strength $\Sigma^{\mathrm{F}}$ on the system size
and the volume fraction.  This model is based on the assumption that
the weakest band bears all the plastic strain and governs the
ultimate yield strength $\Sigma^{\mathrm{F}}$ of the entire system.
The ultimate yield strength $\Sigma^{\mathrm{F}}$ is then given by a
combination of the yield strength of the pure matrix and of the hard
inclusions, weighted respectively by the fraction of matrix sites and
of hard inclusions in the weakest band.  Using extremal statistics
arguments, we proposed an analytical estimate of the average number
of hard inclusions in the weakest band in the limit of large system
sizes. The comparison of the analytical estimate with the simulation
results is satisfactory and our model consequently makes a direct link
between the structure, represented by the plastic threshold on each
site, and the mechanical behavior.

\begin{acknowledgments}
DV acknowledges Anne Canteaut for the indication of Refs.~\cite{Blondeau-PhD11,Blondeau-DCC11}.
CL acknowledges Eric Lu\c con for the help in deriving the exact formula of the average
minimum of hard sites on a diagonal. 
\end{acknowledgments}

\appendix

\section*{Appendix: Analytical approach}
\label{app:estimationMin}

\subsection{Exact formula}
To obtain an exact formula for the average minimum
number $\langle m \rangle$ of inclusions per diagonal,
we use its definition:
\begin{align}
\langle m \rangle &= \sum_{n=0}^{N} n P(m = n)\;,\\
                  &= \sum_{n=1}^{N} n P(m = n)\;.
\end{align}
The variable $n$ in this expression can be reformulated as a sum:
\begin{align}
\langle m \rangle &= \sum_{n=1}^{N} \sum_{k=0}^{n-1} P(m = n)\;,\\
                  &= \sum_{k=0}^{N-1} \sum_{n=k+1}^{N} P(m = n)\;,\\
\end{align}
where the indices in the two sums are enumerated in two different and equivalent ways.
The sum over the index $n$ can then be contracted into:
\begin{align}
\langle m \rangle &= \sum_{k=0}^{N-1} P(m \geq k+1)\;,\\
                  &=\sum_{n=1}^{N} P(m \geq n)\;,\\
                  &= \sum_{n=1}^{N} \Bigl(P(N_d \geq n)\Bigr)^N\;,\\
\label{eq:minExprI}
                  &=  \sum_{n=1}^{N} \Bigl( I_{\phi}(n, N-n+1)\Bigr)^N\;,
\end{align}
where $N_d$ is the random variable counting the
number of hard sites in a diagonal and $I_{\phi}$ the regularized incomplete beta function.
The regularized incomplete beta function is used here to express the cumulative distribution
of the binomial distribution.
The formula Eq.~\ref{eq:minExprI} is explicit but hard to evaluate for large
values of $N$.

\subsection{Analytical estimation}
To obtain an analytical estimate of the average minimum
number $\langle m \rangle$ of inclusions per diagonal,
we use a result from extreme value theory on the minimum
of $N$ independent and identically distributed random variables:
\begin{equation}
\label{eq:extremeValue}
P(N_d \leq \langle m\rangle) = \frac{1}{N}, \quad N \rightarrow \infty\;,
\end{equation}
where $N_d$ is the random variable counting the number of inclusions
in any diagonal.
In our case, the diagonal have $N$ sites.
We introduce the ratio $f = \langle m \rangle/N$ convenient
in the limit of large $N$ and Eq.~\ref{eq:extremeValue} becomes:
\begin{equation}
\label{eq:extremeValuef}
P(N_d \leq f N) = \frac{1}{N}, \quad N \rightarrow \infty\;.
\end{equation}
As the random variable $N_d$ has a binomial distribution 
with parameters $N$ and $\phi$, we can employ
a result of cryptography given in Refs.~\cite{Blondeau-PhD11,Blondeau-DCC11}:
\begin{equation}
\label{eq:blondeau1}
P(N_d \leq f N) =  \frac{\phi \sqrt{1-f}}{(\phi - f)\sqrt{2\pi N f}} e^{-N D(f||\phi)}\;,  
\end{equation}
for $N \rightarrow \infty$ where $D(f||\phi)$ is the Kullback-Leibler
divergence defined as:
\begin{equation}
D(f||\phi) = f \ln{\frac{f}{\phi}} + (1-f)\ln{\frac{1-f}{1-\phi}}\;.
\end{equation}
We then introduce $\epsilon = \phi - f$; and $\epsilon$ is expected to
tend to $0$ in the limit of large $N$, i.e.,  $f$ tends to $\phi$ in
the limit of large $N$.  Using Eqs.~\ref{eq:extremeValuef}
and~\ref{eq:blondeau1} and another result from
Refs.~\cite{Blondeau-PhD11,Blondeau-DCC11} on the behavior of the
Kullback-Leibler divergence valid for $O(\epsilon) = O(\epsilon/\phi)
= O(\epsilon/(1-\phi))$:
\begin{equation}
\label{eq:blondeau2}
D(\phi -\epsilon||\phi) = \frac{\epsilon^2}{2\phi(1-\phi)} + O(\epsilon^3)\;,
\end{equation}
we obtain:
\begin{equation}
\label{eq:epsilon}
\frac{1}{N} = \frac{\phi \sqrt{1-\phi + \epsilon}}{\epsilon\sqrt{2\pi N (\phi -\epsilon)}} e^{-\frac{N\epsilon^2}{2\phi(1-\phi)}}\;.
\end{equation}
To first order in $\epsilon$, this becomes:
\begin{equation}
\label{eq:espilonSimple2}
      \frac{1}{N}      = \frac{1}{\sqrt{2\pi}}\frac{\sqrt{\phi(1-\phi)/N}}{\epsilon}
 e^{-\frac{N\epsilon^2}{2\phi(1-\phi)}}\;,
\end{equation}
We define $\epsilon' = \epsilon/\sqrt{\phi(1-\phi)/N}$ and Eq.~\ref{eq:espilonSimple2} is equivalent to
\begin{equation}
\label{eq:transcendental}
\epsilon'^2 = 2 \ln{\frac{N}{\sqrt{2\pi}\epsilon'}} \;.
\end{equation}
To obtain an approximate solution to this transcendental equation, we  define the variable $r$ such that:
\begin{equation}
\epsilon' = \sqrt{2 h_N} (1 + r)\;,
\end{equation}
where $h_N = \ln{\frac{N}{\sqrt{2\pi}}}$.
The variable $r$ tends to $0$ in the limit of large $N$. 
We also have:
\begin{equation}
\label{eq:espilonSq1}
\epsilon'^2 = 2h_N (1 + r)^2 =  2 h_N (1+ 2r) + {\cal O}(r^2)\;,
\end{equation}
to first order in $r$.  Using the transcendental
equation~\ref{eq:transcendental} and iterating once in $\epsilon$,
we get:
\begin{equation}
\label{eq:espilonSq2}
\epsilon'^2 = 2 \ln{\frac{N}{\sqrt{2\pi} \sqrt{2 h_N} (1 + r) }}\;.
\end{equation}
Equating the right hand sides of Eqs.~\ref{eq:espilonSq1} and~\ref{eq:espilonSq2} leads:
\begin{align}
2 h_N (1+ 2r) &= 2 \ln{\frac{N}{\sqrt{2\pi} \sqrt{2 h_N} (1 + r) }}\;,\\
              &= \ln{\frac{N}{\sqrt{2\pi} \sqrt{2 h_N}}} -2 \ln{(1+r)}\;,\\
\label{eq:r}
              &= \ln{\frac{N}{\sqrt{2\pi} \sqrt{2 h_N}}} -2 r\;.
\end{align}
to first order in $r$.
Eq.~\ref{eq:r} is a linear equation in $r$, its solution reads:
\begin{equation}
 r = \frac{\ln{\frac{N}{\sqrt{2\pi} \sqrt{2 h_N}}} -h_N  }{2h_N +1}\;.
\end{equation}
Finally, the ratio $f = \langle m \rangle/N$ can be expressed in terms of $r(N)$:
\begin{equation}
\label{eq:resultf}
f = \phi - \sqrt{\frac{\phi (1-\phi)}{N}}  \sqrt{2 h_N} (1+ r(N))\;.
\end{equation}

\begin{figure}
  \includegraphics[width=0.9\columnwidth]{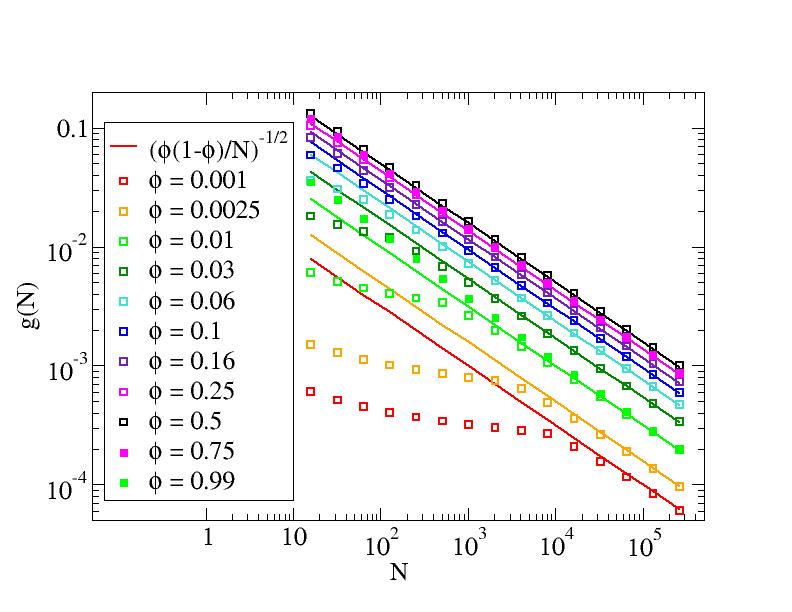}
  \caption{\label{fig:gOfNestimation}
(color online) Variation of the 
function $g(N) = (\phi - f)/(1+ r(N))/\sqrt{2 h_N}$ with $N$
for different values of $\phi$.
The opened squares correspond to a numerical evaluation
of $g(N)$ using $10,000$ iterations of $N$ drawings from a binomial
distribution with parameters $N$ and $\phi$.
The solid and opened squares of the same color 
correspond to results for $\phi$ and $(1-\phi)$, respectively.
The solid lines correspond to the analytical estimation 
given in Eq.~\ref{eq:analyticalEstimation} in the limit
of large $N$.
}
\end{figure}

\begin{figure}
  \includegraphics[width=0.9\columnwidth]{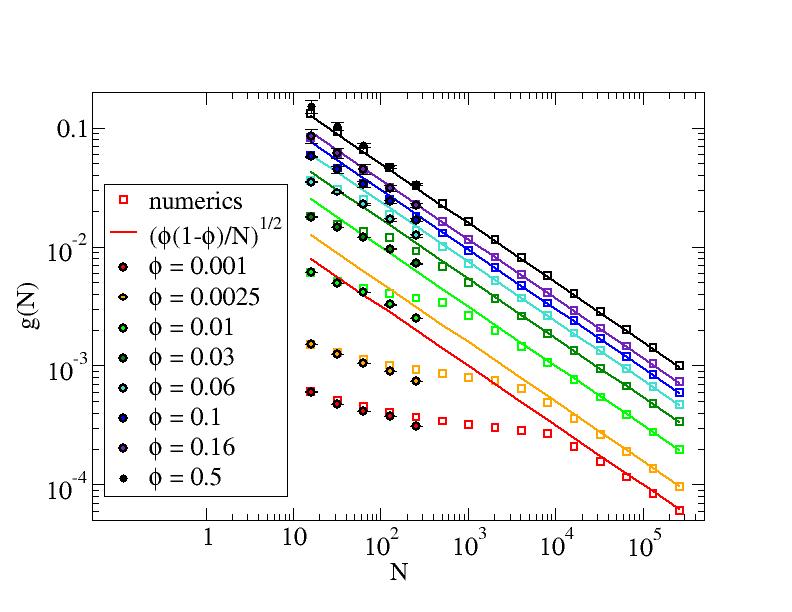}
  \caption{\label{fig:gOfNnumerics}
(color online) Variation of the
function $g(N) = (\phi - f)/(1+ r(N))/\sqrt{2 h_N}$ with $N$
for different values of $\phi$.
The solid circles correspond to simulation results.
The opened squares correspond to a numerical evaluation
of $g(N)$ using $10,000$ iterations of $N$ drawings from a binomial
distribution with parameters $N$ and $\phi$.
The solid lines correspond to the analytical estimation
given in Eq.~\ref{eq:analyticalEstimation} in the limit
of large $N$.
}
\end{figure}

The solution given in Eq.~\ref{eq:resultf} can be checked numerically.
The function $g$ is defined by:
\begin{equation}
g(N) = \frac{\phi - f}{1+ r(N)} \frac{1}{\sqrt{2 h_N}}\;.
\end{equation}
According to Eq.~\ref{eq:resultf}, it is equal to 
\begin{equation}
\label{eq:analyticalEstimation}
g(N) = \sqrt{\frac{\phi (1-\phi)}{N}}\;.
\end{equation}
Fig.~\ref{fig:gOfNestimation} displays the function $g$ as obtained
for $10,000$ numerical iterations of $N$ drawings from a binomial
distribution with parameters $N$ and $\phi$, for different values of $N$ and $\phi$.
The analytical estimation of $g(N)$ given in Eq.~\ref{eq:analyticalEstimation}
works very well down to $N = 16$ for values of $\phi$ in $0.1 \leq \phi \leq 0.9$.
It is expected as we used Eq.~\ref{eq:blondeau2} valid for
$O(\epsilon) = O(\epsilon/\phi) = O(\epsilon/(1-\phi))$. 
For extreme values of $\phi$ the analytical estimation only gives
satisfactory results for very large $N$. The $\phi$-$(1-\phi)$
symmetry is then recovered.
However, the numerical evaluation of the minimum is in very good agreement with
the simulation results for all values of $\phi$ as is shown in Fig.~\ref{fig:gOfNnumerics}.
In this figure, the fraction $f$ of hard sites in the weakest band is estimated
from the simulation results as $(\Sigma^{\mathrm{F}}(\phi,N) -\Sigma^{\mathrm{A}}(N))/(\Sigma^{\mathrm{H}} - \Sigma^{\mathrm{A}}(N))$.

\begin{thebibliography}{51}%
\makeatletter
\providecommand \@ifxundefined [1]{%
 \@ifx{#1\undefined}
}%
\providecommand \@ifnum [1]{%
 \ifnum #1\expandafter \@firstoftwo
 \else \expandafter \@secondoftwo
 \fi
}%
\providecommand \@ifx [1]{%
 \ifx #1\expandafter \@firstoftwo
 \else \expandafter \@secondoftwo
 \fi
}%
\providecommand \natexlab [1]{#1}%
\providecommand \enquote  [1]{``#1''}%
\providecommand \bibnamefont  [1]{#1}%
\providecommand \bibfnamefont [1]{#1}%
\providecommand \citenamefont [1]{#1}%
\providecommand \href@noop [0]{\@secondoftwo}%
\providecommand \href [0]{\begingroup \@sanitize@url \@href}%
\providecommand \@href[1]{\@@startlink{#1}\@@href}%
\providecommand \@@href[1]{\endgroup#1\@@endlink}%
\providecommand \@sanitize@url [0]{\catcode `\\12\catcode `\$12\catcode
  `\&12\catcode `\#12\catcode `\^12\catcode `\_12\catcode `\%12\relax}%
\providecommand \@@startlink[1]{}%
\providecommand \@@endlink[0]{}%
\providecommand \url  [0]{\begingroup\@sanitize@url \@url }%
\providecommand \@url [1]{\endgroup\@href {#1}{\urlprefix }}%
\providecommand \urlprefix  [0]{URL }%
\providecommand \Eprint [0]{\href }%
\providecommand \doibase [0]{http://dx.doi.org/}%
\providecommand \selectlanguage [0]{\@gobble}%
\providecommand \bibinfo  [0]{\@secondoftwo}%
\providecommand \bibfield  [0]{\@secondoftwo}%
\providecommand \translation [1]{[#1]}%
\providecommand \BibitemOpen [0]{}%
\providecommand \bibitemStop [0]{}%
\providecommand \bibitemNoStop [0]{.\EOS\space}%
\providecommand \EOS [0]{\spacefactor3000\relax}%
\providecommand \BibitemShut  [1]{\csname bibitem#1\endcsname}%
\let\auto@bib@innerbib\@empty
\bibitem [{\citenamefont {Torquato}(2002)}]{Torquato-book02}%
  \BibitemOpen
  \bibfield  {author} {\bibinfo {author} {\bibfnamefont {S.}~\bibnamefont
  {Torquato}},\ }\href@noop {} {\emph {\bibinfo {title} {Random Heterogeneous
  Materials. Microstructure and Macroscopic Properties}}}\ (\bibinfo
  {publisher} {Springer, New York},\ \bibinfo {year} {2002})\BibitemShut
  {NoStop}%
\bibitem [{\citenamefont {Branthaver}\ \emph {et~al.}(1993)\citenamefont
  {Branthaver}, \citenamefont {Pedersen}, \citenamefont {Robertson},
  \citenamefont {Duvall}, \citenamefont {Kim}, \citenamefont {Harnsberger},
  \citenamefont {Mill}, \citenamefont {Ensley}, \citenamefont {Barbour},\ and\
  \citenamefont {Schabron}}]{Branthaver-SHRP93}%
  \BibitemOpen
  \bibfield  {author} {\bibinfo {author} {\bibfnamefont {J.~F.}\ \bibnamefont
  {Branthaver}}, \bibinfo {author} {\bibfnamefont {J.~C.}\ \bibnamefont
  {Pedersen}}, \bibinfo {author} {\bibfnamefont {R.~E.}\ \bibnamefont
  {Robertson}}, \bibinfo {author} {\bibfnamefont {J.~J.}\ \bibnamefont
  {Duvall}}, \bibinfo {author} {\bibfnamefont {S.~S.}\ \bibnamefont {Kim}},
  \bibinfo {author} {\bibfnamefont {P.~M.}\ \bibnamefont {Harnsberger}},
  \bibinfo {author} {\bibfnamefont {T.}~\bibnamefont {Mill}}, \bibinfo {author}
  {\bibfnamefont {E.~K.}\ \bibnamefont {Ensley}}, \bibinfo {author}
  {\bibfnamefont {F.~A.}\ \bibnamefont {Barbour}}, \ and\ \bibinfo {author}
  {\bibfnamefont {J.~F.}\ \bibnamefont {Schabron}},\ }\href@noop {} {\emph
  {\bibinfo {title} {Binder Characterization and Evaluation. Volume 2:
  Chemistry}}},\ \bibinfo {type} {Tech. Rep.}\ \bibinfo {number} {SHRP-A-368}\
  (\bibinfo  {institution} {Strategic Highway Research Program},\ \bibinfo
  {year} {1993})\BibitemShut {NoStop}%
\bibitem [{\citenamefont {Anderson}\ \emph {et~al.}(1994)\citenamefont
  {Anderson}, \citenamefont {Christensen}, \citenamefont {Bahia}, \citenamefont
  {Dongre}, \citenamefont {Sharma}, \citenamefont {Antle},\ and\ \citenamefont
  {Button}}]{Anderson-SHRP94}%
  \BibitemOpen
  \bibfield  {author} {\bibinfo {author} {\bibfnamefont {D.~A.}\ \bibnamefont
  {Anderson}}, \bibinfo {author} {\bibfnamefont {D.~W.}\ \bibnamefont
  {Christensen}}, \bibinfo {author} {\bibfnamefont {H.~U.}\ \bibnamefont
  {Bahia}}, \bibinfo {author} {\bibfnamefont {R.}~\bibnamefont {Dongre}},
  \bibinfo {author} {\bibfnamefont {M.~G.}\ \bibnamefont {Sharma}}, \bibinfo
  {author} {\bibfnamefont {C.~E.}\ \bibnamefont {Antle}}, \ and\ \bibinfo
  {author} {\bibfnamefont {J.}~\bibnamefont {Button}},\ }\href@noop {} {\emph
  {\bibinfo {title} {Binder Characterization and Evaluation. Volume 3: Physical
  Characterization}}},\ \bibinfo {type} {Tech. Rep.}\ \bibinfo {number}
  {A-369}\ (\bibinfo  {institution} {Strategic Highway Research Program},\
  \bibinfo {year} {1994})\BibitemShut {NoStop}%
\bibitem [{\citenamefont {Chen}\ and\ \citenamefont
  {Peng}(1998)}]{Chen-JMCE98}%
  \BibitemOpen
  \bibfield  {author} {\bibinfo {author} {\bibfnamefont {J.-S.}\ \bibnamefont
  {Chen}}\ and\ \bibinfo {author} {\bibfnamefont {C.-H.}\ \bibnamefont
  {Peng}},\ }\href@noop {} {\bibfield  {journal} {\bibinfo  {journal} {J.
  Mater. Civ. Eng.}\ }\textbf {\bibinfo {volume} {10}},\ \bibinfo {pages} {256}
  (\bibinfo {year} {1998})}\BibitemShut {NoStop}%
\bibitem [{\citenamefont {You}\ \emph {et~al.}(2012)\citenamefont {You},
  \citenamefont {Abu Al-Rub}, \citenamefont {Darabi}, \citenamefont {Masad},\
  and\ \citenamefont {Little}}]{You-CBM12}%
  \BibitemOpen
  \bibfield  {author} {\bibinfo {author} {\bibfnamefont {T.}~\bibnamefont
  {You}}, \bibinfo {author} {\bibfnamefont {R.~K.}\ \bibnamefont {Abu Al-Rub}},
  \bibinfo {author} {\bibfnamefont {M.~K.}\ \bibnamefont {Darabi}}, \bibinfo
  {author} {\bibfnamefont {E.~A.}\ \bibnamefont {Masad}}, \ and\ \bibinfo
  {author} {\bibfnamefont {D.~N.}\ \bibnamefont {Little}},\ }\href@noop {}
  {\bibfield  {journal} {\bibinfo  {journal} {Constr. Build. Mater.}\ }\textbf
  {\bibinfo {volume} {28}},\ \bibinfo {pages} {531} (\bibinfo {year}
  {2012})}\BibitemShut {NoStop}%
\bibitem [{\citenamefont {Hofmann}\ \emph {et~al.}(2008)\citenamefont
  {Hofmann}, \citenamefont {Suh}, \citenamefont {Wiest}, \citenamefont {Duan},
  \citenamefont {Lind}, \citenamefont {Demetriou},\ and\ \citenamefont
  {Johnson}}]{Hofmann-Nat08}%
  \BibitemOpen
  \bibfield  {author} {\bibinfo {author} {\bibfnamefont {D.~C.}\ \bibnamefont
  {Hofmann}}, \bibinfo {author} {\bibfnamefont {J.-Y.}\ \bibnamefont {Suh}},
  \bibinfo {author} {\bibfnamefont {A.}~\bibnamefont {Wiest}}, \bibinfo
  {author} {\bibfnamefont {G.}~\bibnamefont {Duan}}, \bibinfo {author}
  {\bibfnamefont {M.-L.}\ \bibnamefont {Lind}}, \bibinfo {author}
  {\bibfnamefont {M.~D.}\ \bibnamefont {Demetriou}}, \ and\ \bibinfo {author}
  {\bibfnamefont {W.~L.}\ \bibnamefont {Johnson}},\ }\href@noop {} {\bibfield
  {journal} {\bibinfo  {journal} {Nature}\ }\textbf {\bibinfo {volume} {451}},\
  \bibinfo {pages} {1085} (\bibinfo {year} {2008})}\BibitemShut {NoStop}%
\bibitem [{\citenamefont {Ferry}\ \emph {et~al.}(2013)\citenamefont {Ferry},
  \citenamefont {Laws}, \citenamefont {White}, \citenamefont {Miskovic},
  \citenamefont {Shamlaye}, \citenamefont {Xu},\ and\ \citenamefont
  {Biletska}}]{Ferry-MRS13}%
  \BibitemOpen
  \bibfield  {author} {\bibinfo {author} {\bibfnamefont {M.}~\bibnamefont
  {Ferry}}, \bibinfo {author} {\bibfnamefont {K.~J.}\ \bibnamefont {Laws}},
  \bibinfo {author} {\bibfnamefont {C.}~\bibnamefont {White}}, \bibinfo
  {author} {\bibfnamefont {D.~M.}\ \bibnamefont {Miskovic}}, \bibinfo {author}
  {\bibfnamefont {K.~F.}\ \bibnamefont {Shamlaye}}, \bibinfo {author}
  {\bibfnamefont {W.}~\bibnamefont {Xu}}, \ and\ \bibinfo {author}
  {\bibfnamefont {O.}~\bibnamefont {Biletska}},\ }\href@noop {} {\bibfield
  {journal} {\bibinfo  {journal} {MRS Comm.}\ }\textbf {\bibinfo {volume}
  {3}},\ \bibinfo {pages} {1} (\bibinfo {year} {2013})}\BibitemShut {NoStop}%
\bibitem [{\citenamefont {Dasgupta}\ \emph {et~al.}(2013)\citenamefont
  {Dasgupta}, \citenamefont {Mishra}, \citenamefont {Procaccia},\ and\
  \citenamefont {Samwer}}]{Samwer-APL13}%
  \BibitemOpen
  \bibfield  {author} {\bibinfo {author} {\bibfnamefont {R.}~\bibnamefont
  {Dasgupta}}, \bibinfo {author} {\bibfnamefont {P.}~\bibnamefont {Mishra}},
  \bibinfo {author} {\bibfnamefont {I.}~\bibnamefont {Procaccia}}, \ and\
  \bibinfo {author} {\bibfnamefont {K.}~\bibnamefont {Samwer}},\ }\href
  {\doibase {10.1063/1.4805033}} {\bibfield  {journal} {\bibinfo  {journal}
  {Appl. Phys. Lett.}\ }\textbf {\bibinfo {volume} {{102}}} (\bibinfo {year}
  {{2013}}),\ {10.1063/1.4805033}}\BibitemShut {NoStop}%
\bibitem [{\citenamefont {Gendelman}\ \emph {et~al.}(2014)\citenamefont
  {Gendelman}, \citenamefont {Joy}, \citenamefont {Mishra}, \citenamefont
  {Procaccia},\ and\ \citenamefont {Samwer}}]{Samwer-ActaMat14}%
  \BibitemOpen
  \bibfield  {author} {\bibinfo {author} {\bibfnamefont {O.}~\bibnamefont
  {Gendelman}}, \bibinfo {author} {\bibfnamefont {A.}~\bibnamefont {Joy}},
  \bibinfo {author} {\bibfnamefont {P.}~\bibnamefont {Mishra}}, \bibinfo
  {author} {\bibfnamefont {I.}~\bibnamefont {Procaccia}}, \ and\ \bibinfo
  {author} {\bibfnamefont {K.}~\bibnamefont {Samwer}},\ }\href {\doibase
  10.1016/j.actamat.2013.10.029} {\bibfield  {journal} {\bibinfo  {journal}
  {Acta Mater.}\ }\textbf {\bibinfo {volume} {63}},\ \bibinfo {pages} {209}
  (\bibinfo {year} {2014})}\BibitemShut {NoStop}%
\bibitem [{\citenamefont {Abu Al-Rub}(2009)}]{Al-Rub09}%
  \BibitemOpen
  \bibfield  {author} {\bibinfo {author} {\bibfnamefont {R.}~\bibnamefont {Abu
  Al-Rub}},\ }\href@noop {} {\bibfield  {journal} {\bibinfo  {journal} {Int. J.
  Multiscale Comp. Engin.}\ }\textbf {\bibinfo {volume} {7}} (\bibinfo {year}
  {2009})}\BibitemShut {NoStop}%
\bibitem [{\citenamefont {Abu Al-Rub}\ \emph {et~al.}(2011)\citenamefont {Abu
  Al-Rub}, \citenamefont {Darabi}, \citenamefont {You}, \citenamefont {Masad},\
  and\ \citenamefont {Little}}]{Al-Rub11}%
  \BibitemOpen
  \bibfield  {author} {\bibinfo {author} {\bibfnamefont {R.~K.}\ \bibnamefont
  {Abu Al-Rub}}, \bibinfo {author} {\bibfnamefont {M.~K.}\ \bibnamefont
  {Darabi}}, \bibinfo {author} {\bibfnamefont {T.}~\bibnamefont {You}},
  \bibinfo {author} {\bibfnamefont {E.~A.}\ \bibnamefont {Masad}}, \ and\
  \bibinfo {author} {\bibfnamefont {D.~N.}\ \bibnamefont {Little}},\
  }\href@noop {} {\bibfield  {journal} {\bibinfo  {journal} {Int. J. Roads
  Airports}\ }\textbf {\bibinfo {volume} {1}},\ \bibinfo {pages} {1} (\bibinfo
  {year} {2011})}\BibitemShut {NoStop}%
\bibitem [{\citenamefont {Abu Al-Rub}\ \emph {et~al.}(2009)\citenamefont {Abu
  Al-Rub}, \citenamefont {You}, \citenamefont {Masad},\ and\ \citenamefont
  {Little}}]{Al-Rub12}%
  \BibitemOpen
  \bibfield  {author} {\bibinfo {author} {\bibfnamefont {R.~K.}\ \bibnamefont
  {Abu Al-Rub}}, \bibinfo {author} {\bibfnamefont {T.}~\bibnamefont {You}},
  \bibinfo {author} {\bibfnamefont {E.~A.}\ \bibnamefont {Masad}}, \ and\
  \bibinfo {author} {\bibfnamefont {D.~N.}\ \bibnamefont {Little}},\
  }\href@noop {} {\bibfield  {journal} {\bibinfo  {journal} {Int. J. Adv. Eng.
  Sci. Appl. Math.}\ }\textbf {\bibinfo {volume} {7}},\ \bibinfo {pages} {14}
  (\bibinfo {year} {2009})}\BibitemShut {NoStop}%
\bibitem [{\citenamefont {Roux}\ \emph {et~al.}(2003)\citenamefont {Roux},
  \citenamefont {Vandembroucq},\ and\ \citenamefont {Hild}}]{RVH-EJMA03}%
  \BibitemOpen
  \bibfield  {author} {\bibinfo {author} {\bibfnamefont {S.}~\bibnamefont
  {Roux}}, \bibinfo {author} {\bibfnamefont {D.}~\bibnamefont {Vandembroucq}},
  \ and\ \bibinfo {author} {\bibfnamefont {F.}~\bibnamefont {Hild}},\
  }\href@noop {} {\bibfield  {journal} {\bibinfo  {journal} {Eur. J. Mech. A}\
  }\textbf {\bibinfo {volume} {22}},\ \bibinfo {pages} {743} (\bibinfo {year}
  {2003})}\BibitemShut {NoStop}%
\bibitem [{\citenamefont {Patinet}\ \emph {et~al.}(2013)\citenamefont
  {Patinet}, \citenamefont {Vandembroucq},\ and\ \citenamefont
  {Roux}}]{PVR-PRL13}%
  \BibitemOpen
  \bibfield  {author} {\bibinfo {author} {\bibfnamefont {S.}~\bibnamefont
  {Patinet}}, \bibinfo {author} {\bibfnamefont {D.}~\bibnamefont
  {Vandembroucq}}, \ and\ \bibinfo {author} {\bibfnamefont {S.}~\bibnamefont
  {Roux}},\ }\href@noop {} {\bibfield  {journal} {\bibinfo  {journal} {Phys.
  Rev. Lett.}\ }\textbf {\bibinfo {volume} {110}},\ \bibinfo {pages} {165507}
  (\bibinfo {year} {2013})}\BibitemShut {NoStop}%
\bibitem [{\citenamefont {Ponte-Castaneda}\ and\ \citenamefont
  {DeBotton}(1992)}]{Ponte-Castaneda-PRSA92}%
  \BibitemOpen
  \bibfield  {author} {\bibinfo {author} {\bibfnamefont {P.}~\bibnamefont
  {Ponte-Castaneda}}\ and\ \bibinfo {author} {\bibfnamefont {G.}~\bibnamefont
  {DeBotton}},\ }\href@noop {} {\bibfield  {journal} {\bibinfo  {journal}
  {Proc. Roy. Soc. A}\ }\textbf {\bibinfo {volume} {438}},\ \bibinfo {pages}
  {419} (\bibinfo {year} {1992})}\BibitemShut {NoStop}%
\bibitem [{\citenamefont {DeBotton}(1995)}]{Debotton-IJSS95}%
  \BibitemOpen
  \bibfield  {author} {\bibinfo {author} {\bibfnamefont {G.}~\bibnamefont
  {DeBotton}},\ }\href@noop {} {\bibfield  {journal} {\bibinfo  {journal} {Int.
  J. Solids Struct.}\ }\textbf {\bibinfo {volume} {32}},\ \bibinfo {pages}
  {1743} (\bibinfo {year} {1995})}\BibitemShut {NoStop}%
\bibitem [{\citenamefont {Fleck}\ and\ \citenamefont
  {Willis}(2004)}]{Willis-JMPS04}%
  \BibitemOpen
  \bibfield  {author} {\bibinfo {author} {\bibfnamefont {N.~A.}\ \bibnamefont
  {Fleck}}\ and\ \bibinfo {author} {\bibfnamefont {J.~R.}\ \bibnamefont
  {Willis}},\ }\href@noop {} {\bibfield  {journal} {\bibinfo  {journal} {J.
  Mech. Phys. Solids}\ }\textbf {\bibinfo {volume} {52}},\ \bibinfo {pages}
  {1855} (\bibinfo {year} {2004})}\BibitemShut {NoStop}%
\bibitem [{\citenamefont {Turgeman}\ and\ \citenamefont
  {Guessab}(2011)}]{Turgeman-MRC11}%
  \BibitemOpen
  \bibfield  {author} {\bibinfo {author} {\bibfnamefont {S.}~\bibnamefont
  {Turgeman}}\ and\ \bibinfo {author} {\bibfnamefont {B.}~\bibnamefont
  {Guessab}},\ }\href@noop {} {\bibfield  {journal} {\bibinfo  {journal} {Mech.
  Res. Comm.}\ }\textbf {\bibinfo {volume} {38}},\ \bibinfo {pages} {181}
  (\bibinfo {year} {2011})}\BibitemShut {NoStop}%
\bibitem [{\citenamefont {Suquet}\ and\ \citenamefont
  {Lahellec}(2014)}]{Suquet-IUTAM14}%
  \BibitemOpen
  \bibfield  {author} {\bibinfo {author} {\bibfnamefont {P.}~\bibnamefont
  {Suquet}}\ and\ \bibinfo {author} {\bibfnamefont {N.}~\bibnamefont
  {Lahellec}},\ }\href@noop {} {\bibfield  {journal} {\bibinfo  {journal}
  {Procedia IUTAM}\ }\textbf {\bibinfo {volume} {10}},\ \bibinfo {pages} {247}
  (\bibinfo {year} {2014})}\BibitemShut {NoStop}%
\bibitem [{\citenamefont {Bulatov}\ and\ \citenamefont
  {Argon}(1994)}]{BulatovArgon94a}%
  \BibitemOpen
  \bibfield  {author} {\bibinfo {author} {\bibfnamefont {V.~V.}\ \bibnamefont
  {Bulatov}}\ and\ \bibinfo {author} {\bibfnamefont {A.~S.}\ \bibnamefont
  {Argon}},\ }\href@noop {} {\bibfield  {journal} {\bibinfo  {journal} {Modell.
  Simul. Mater. Sci. Eng.}\ }\textbf {\bibinfo {volume} {2}},\ \bibinfo {pages}
  {167} (\bibinfo {year} {1994})}\BibitemShut {NoStop}%
\bibitem [{\citenamefont {Baret}\ \emph {et~al.}(2002)\citenamefont {Baret},
  \citenamefont {Vandembroucq},\ and\ \citenamefont {Roux}}]{BVR-PRL02}%
  \BibitemOpen
  \bibfield  {author} {\bibinfo {author} {\bibfnamefont {J.-C.}\ \bibnamefont
  {Baret}}, \bibinfo {author} {\bibfnamefont {D.}~\bibnamefont {Vandembroucq}},
  \ and\ \bibinfo {author} {\bibfnamefont {S.}~\bibnamefont {Roux}},\
  }\href@noop {} {\bibfield  {journal} {\bibinfo  {journal} {Phys. Rev. Lett.}\
  }\textbf {\bibinfo {volume} {89}},\ \bibinfo {pages} {195506} (\bibinfo
  {year} {2002})}\BibitemShut {NoStop}%
\bibitem [{\citenamefont {Picard}\ \emph {et~al.}(2002)\citenamefont {Picard},
  \citenamefont {Ajdari}, \citenamefont {Bocquet},\ and\ \citenamefont
  {Lequeux}}]{Picard-PRE02}%
  \BibitemOpen
  \bibfield  {author} {\bibinfo {author} {\bibfnamefont {G.}~\bibnamefont
  {Picard}}, \bibinfo {author} {\bibfnamefont {A.}~\bibnamefont {Ajdari}},
  \bibinfo {author} {\bibfnamefont {L.}~\bibnamefont {Bocquet}}, \ and\
  \bibinfo {author} {\bibfnamefont {F.}~\bibnamefont {Lequeux}},\ }\href@noop
  {} {\bibfield  {journal} {\bibinfo  {journal} {Phys. Rev. E}\ }\textbf
  {\bibinfo {volume} {66}},\ \bibinfo {pages} {051501} (\bibinfo {year}
  {2002})}\BibitemShut {NoStop}%
\bibitem [{\citenamefont {Homer}\ and\ \citenamefont
  {Schuh}(2009)}]{Schuh-ActaMat09}%
  \BibitemOpen
  \bibfield  {author} {\bibinfo {author} {\bibfnamefont {E.~R.}\ \bibnamefont
  {Homer}}\ and\ \bibinfo {author} {\bibfnamefont {C.~A.}\ \bibnamefont
  {Schuh}},\ }\href@noop {} {\bibfield  {journal} {\bibinfo  {journal} {Acta
  Mat.}\ }\textbf {\bibinfo {volume} {57}},\ \bibinfo {pages} {2823} (\bibinfo
  {year} {2009})}\BibitemShut {NoStop}%
\bibitem [{\citenamefont {Talamali}\ \emph {et~al.}(2012)\citenamefont
  {Talamali}, \citenamefont {Pet\"aj\"a}, \citenamefont {Vandembroucq},\ and\
  \citenamefont {Roux}}]{TPVR-CRM12}%
  \BibitemOpen
  \bibfield  {author} {\bibinfo {author} {\bibfnamefont {M.}~\bibnamefont
  {Talamali}}, \bibinfo {author} {\bibfnamefont {V.}~\bibnamefont
  {Pet\"aj\"a}}, \bibinfo {author} {\bibfnamefont {D.}~\bibnamefont
  {Vandembroucq}}, \ and\ \bibinfo {author} {\bibfnamefont {S.}~\bibnamefont
  {Roux}},\ }\href@noop {} {\bibfield  {journal} {\bibinfo  {journal} {C.R.
  M\'ecanique}\ }\textbf {\bibinfo {volume} {340}},\ \bibinfo {pages} {275}
  (\bibinfo {year} {2012})}\BibitemShut {NoStop}%
\bibitem [{\citenamefont {Nicolas}\ \emph {et~al.}(2014)\citenamefont
  {Nicolas}, \citenamefont {Martens}, \citenamefont {Bocquet},\ and\
  \citenamefont {Barrat}}]{Nicolas-SM14}%
  \BibitemOpen
  \bibfield  {author} {\bibinfo {author} {\bibfnamefont {A.}~\bibnamefont
  {Nicolas}}, \bibinfo {author} {\bibfnamefont {K.}~\bibnamefont {Martens}},
  \bibinfo {author} {\bibfnamefont {L.}~\bibnamefont {Bocquet}}, \ and\
  \bibinfo {author} {\bibfnamefont {J.-L.}\ \bibnamefont {Barrat}},\
  }\href@noop {} {\bibfield  {journal} {\bibinfo  {journal} {Soft Matter}\
  }\textbf {\bibinfo {volume} {10}},\ \bibinfo {pages} {4648} (\bibinfo {year}
  {2014})}\BibitemShut {NoStop}%
\bibitem [{\citenamefont {Rodney}\ \emph {et~al.}(2011)\citenamefont {Rodney},
  \citenamefont {Tanguy},\ and\ \citenamefont {Vandembroucq}}]{RTV-MSMSE11}%
  \BibitemOpen
  \bibfield  {author} {\bibinfo {author} {\bibfnamefont {D.}~\bibnamefont
  {Rodney}}, \bibinfo {author} {\bibfnamefont {A.}~\bibnamefont {Tanguy}}, \
  and\ \bibinfo {author} {\bibfnamefont {D.}~\bibnamefont {Vandembroucq}},\
  }\href@noop {} {\bibfield  {journal} {\bibinfo  {journal} {Modelling Simul.
  Mater. Sci. Eng.}\ }\textbf {\bibinfo {volume} {19}},\ \bibinfo {pages}
  {083001} (\bibinfo {year} {2011})}\BibitemShut {NoStop}%
\bibitem [{\citenamefont {Argon}(1979)}]{Argon-ActaMet79}%
  \BibitemOpen
  \bibfield  {author} {\bibinfo {author} {\bibfnamefont {A.~S.}\ \bibnamefont
  {Argon}},\ }\href@noop {} {\bibfield  {journal} {\bibinfo  {journal} {Acta
  Metall.}\ }\textbf {\bibinfo {volume} {27}},\ \bibinfo {pages} {47} (\bibinfo
  {year} {1979})}\BibitemShut {NoStop}%
\bibitem [{\citenamefont {Falk}\ and\ \citenamefont
  {Langer}(1998)}]{FalkLanger-PRE98}%
  \BibitemOpen
  \bibfield  {author} {\bibinfo {author} {\bibfnamefont {M.~L.}\ \bibnamefont
  {Falk}}\ and\ \bibinfo {author} {\bibfnamefont {J.~S.}\ \bibnamefont
  {Langer}},\ }\href@noop {} {\bibfield  {journal} {\bibinfo  {journal} {Phys.
  Rev. E}\ }\textbf {\bibinfo {volume} {57}},\ \bibinfo {pages} {7192}
  (\bibinfo {year} {1998})}\BibitemShut {NoStop}%
\bibitem [{\citenamefont {Eshelby}(1957)}]{Eshelby57}%
  \BibitemOpen
  \bibfield  {author} {\bibinfo {author} {\bibfnamefont {J.~D.}\ \bibnamefont
  {Eshelby}},\ }\href@noop {} {\bibfield  {journal} {\bibinfo  {journal} {Proc.
  Roy. Soc. A}\ }\textbf {\bibinfo {volume} {241}},\ \bibinfo {pages} {376}
  (\bibinfo {year} {1957})}\BibitemShut {NoStop}%
\bibitem [{\citenamefont {Vandembroucq}\ and\ \citenamefont
  {Roux}(2011)}]{VR-PRB11}%
  \BibitemOpen
  \bibfield  {author} {\bibinfo {author} {\bibfnamefont {D.}~\bibnamefont
  {Vandembroucq}}\ and\ \bibinfo {author} {\bibfnamefont {S.}~\bibnamefont
  {Roux}},\ }\href@noop {} {\bibfield  {journal} {\bibinfo  {journal} {Phys.
  Rev. B}\ }\textbf {\bibinfo {volume} {84}},\ \bibinfo {pages} {134210}
  (\bibinfo {year} {2011})}\BibitemShut {NoStop}%
\bibitem [{\citenamefont {Homer}(2014)}]{Homer-ActaMat14}%
  \BibitemOpen
  \bibfield  {author} {\bibinfo {author} {\bibfnamefont {E.~R.}\ \bibnamefont
  {Homer}},\ }\href@noop {} {\bibfield  {journal} {\bibinfo  {journal} {Acta
  Mater.}\ }\textbf {\bibinfo {volume} {63}},\ \bibinfo {pages} {44} (\bibinfo
  {year} {2014})}\BibitemShut {NoStop}%
\bibitem [{\citenamefont {Homer}(2015)}]{Homer-ActaMat15}%
  \BibitemOpen
  \bibfield  {author} {\bibinfo {author} {\bibfnamefont {E.~R.}\ \bibnamefont
  {Homer}},\ }\href@noop {} {\bibfield  {journal} {\bibinfo  {journal} {Acta
  Mater.}\ }\textbf {\bibinfo {volume} {83}},\ \bibinfo {pages} {203} (\bibinfo
  {year} {2015})}\BibitemShut {NoStop}%
\bibitem [{\citenamefont {Talamali}\ \emph {et~al.}(2011)\citenamefont
  {Talamali}, \citenamefont {Pet\"aj\"a}, \citenamefont {Vandembroucq},\ and\
  \citenamefont {Roux}}]{TPVR-PRE11}%
  \BibitemOpen
  \bibfield  {author} {\bibinfo {author} {\bibfnamefont {M.}~\bibnamefont
  {Talamali}}, \bibinfo {author} {\bibfnamefont {V.}~\bibnamefont
  {Pet\"aj\"a}}, \bibinfo {author} {\bibfnamefont {D.}~\bibnamefont
  {Vandembroucq}}, \ and\ \bibinfo {author} {\bibfnamefont {S.}~\bibnamefont
  {Roux}},\ }\href@noop {} {\bibfield  {journal} {\bibinfo  {journal} {Phys.
  Rev. E}\ }\textbf {\bibinfo {volume} {84}},\ \bibinfo {pages} {016115}
  (\bibinfo {year} {2011})}\BibitemShut {NoStop}%
\bibitem [{\citenamefont {Spaepen}(1977)}]{Spaepen-ActaMet77}%
  \BibitemOpen
  \bibfield  {author} {\bibinfo {author} {\bibfnamefont {F.}~\bibnamefont
  {Spaepen}},\ }\href@noop {} {\bibfield  {journal} {\bibinfo  {journal} {Acta
  Metall.}\ }\textbf {\bibinfo {volume} {25}},\ \bibinfo {pages} {407}
  (\bibinfo {year} {1977})}\BibitemShut {NoStop}%
\bibitem [{\citenamefont {Maloney}\ and\ \citenamefont
  {Lema\^{\i}tre}(2004{\natexlab{a}})}]{Maloney-PRL04b}%
  \BibitemOpen
  \bibfield  {author} {\bibinfo {author} {\bibfnamefont {C.~E.}\ \bibnamefont
  {Maloney}}\ and\ \bibinfo {author} {\bibfnamefont {A.}~\bibnamefont
  {Lema\^{\i}tre}},\ }\href@noop {} {\bibfield  {journal} {\bibinfo  {journal}
  {Phys. Rev. Lett.}\ }\textbf {\bibinfo {volume} {93}},\ \bibinfo {pages}
  {195501} (\bibinfo {year} {2004}{\natexlab{a}})}\BibitemShut {NoStop}%
\bibitem [{\citenamefont {Tanguy}\ \emph {et~al.}(2006)\citenamefont {Tanguy},
  \citenamefont {Leonforte},\ and\ \citenamefont {Barrat}}]{Tanguy-EPJE06}%
  \BibitemOpen
  \bibfield  {author} {\bibinfo {author} {\bibfnamefont {A.}~\bibnamefont
  {Tanguy}}, \bibinfo {author} {\bibfnamefont {F.}~\bibnamefont {Leonforte}}, \
  and\ \bibinfo {author} {\bibfnamefont {J.-L.}\ \bibnamefont {Barrat}},\
  }\href@noop {} {\bibfield  {journal} {\bibinfo  {journal} {Eur. Phys. J. E}\
  }\textbf {\bibinfo {volume} {20}},\ \bibinfo {pages} {355} (\bibinfo {year}
  {2006})}\BibitemShut {NoStop}%
\bibitem [{\citenamefont {Lin}\ \emph {et~al.}(2014)\citenamefont {Lin},
  \citenamefont {Lerner}, \citenamefont {Rosso},\ and\ \citenamefont
  {Wyart}}]{Wyart-PNAS14}%
  \BibitemOpen
  \bibfield  {author} {\bibinfo {author} {\bibfnamefont {J.}~\bibnamefont
  {Lin}}, \bibinfo {author} {\bibfnamefont {E.}~\bibnamefont {Lerner}},
  \bibinfo {author} {\bibfnamefont {A.}~\bibnamefont {Rosso}}, \ and\ \bibinfo
  {author} {\bibfnamefont {M.}~\bibnamefont {Wyart}},\ }\href@noop {}
  {\bibfield  {journal} {\bibinfo  {journal} {Proc. Nat. Acd. Sci.}\ }\textbf
  {\bibinfo {volume} {111}},\ \bibinfo {pages} {14382} (\bibinfo {year}
  {2014})}\BibitemShut {NoStop}%
\bibitem [{\citenamefont {Budrikis}\ and\ \citenamefont
  {Zapperi}(2013)}]{Budrikis-PRE13}%
  \BibitemOpen
  \bibfield  {author} {\bibinfo {author} {\bibfnamefont {Z.}~\bibnamefont
  {Budrikis}}\ and\ \bibinfo {author} {\bibfnamefont {S.}~\bibnamefont
  {Zapperi}},\ }\href@noop {} {\bibfield  {journal} {\bibinfo  {journal} {Phys.
  Rev. E}\ }\textbf {\bibinfo {volume} {88}},\ \bibinfo {pages} {062403}
  (\bibinfo {year} {2013})}\BibitemShut {NoStop}%
\bibitem [{\citenamefont {Tyukodi}\ \emph {et~al.}(2015)\citenamefont
  {Tyukodi}, \citenamefont {Patinet}, \citenamefont {Roux},\ and\ \citenamefont
  {Vandembroucq}}]{TPRV-preprint15}%
  \BibitemOpen
  \bibfield  {author} {\bibinfo {author} {\bibfnamefont {B.}~\bibnamefont
  {Tyukodi}}, \bibinfo {author} {\bibfnamefont {S.}~\bibnamefont {Patinet}},
  \bibinfo {author} {\bibfnamefont {S.}~\bibnamefont {Roux}}, \ and\ \bibinfo
  {author} {\bibfnamefont {D.}~\bibnamefont {Vandembroucq}},\ }\href@noop {}
  {\bibfield  {journal} {\bibinfo  {journal} {unpublished}\ } (\bibinfo {year}
  {2015})}\BibitemShut {NoStop}%
\bibitem [{\citenamefont {Maloney}\ and\ \citenamefont
  {Lema\^{\i}tre}(2004{\natexlab{b}})}]{Maloney-PRL04a}%
  \BibitemOpen
  \bibfield  {author} {\bibinfo {author} {\bibfnamefont {C.~E.}\ \bibnamefont
  {Maloney}}\ and\ \bibinfo {author} {\bibfnamefont {A.}~\bibnamefont
  {Lema\^{\i}tre}},\ }\href@noop {} {\bibfield  {journal} {\bibinfo  {journal}
  {Phys. Rev. Lett.}\ }\textbf {\bibinfo {volume} {93}},\ \bibinfo {pages}
  {016001} (\bibinfo {year} {2004}{\natexlab{b}})}\BibitemShut {NoStop}%
\bibitem [{\citenamefont {Maloney}\ and\ \citenamefont
  {Lema\^{\i}tre}(2006)}]{Maloney-PRE06}%
  \BibitemOpen
  \bibfield  {author} {\bibinfo {author} {\bibfnamefont {C.~E.}\ \bibnamefont
  {Maloney}}\ and\ \bibinfo {author} {\bibfnamefont {A.}~\bibnamefont
  {Lema\^{\i}tre}},\ }\href@noop {} {\bibfield  {journal} {\bibinfo  {journal}
  {Phys. Rev. E}\ }\textbf {\bibinfo {volume} {74}},\ \bibinfo {pages} {016118}
  (\bibinfo {year} {2006})}\BibitemShut {NoStop}%
\bibitem [{\citenamefont {Leidner}\ and\ \citenamefont
  {Woodmans}(1974)}]{Leidner-JAPS74}%
  \BibitemOpen
  \bibfield  {author} {\bibinfo {author} {\bibfnamefont {J.}~\bibnamefont
  {Leidner}}\ and\ \bibinfo {author} {\bibfnamefont {R.~T.}\ \bibnamefont
  {Woodmans}},\ }\href@noop {} {\bibfield  {journal} {\bibinfo  {journal} {J.
  Appl. Polymer Sci.}\ }\textbf {\bibinfo {volume} {18}},\ \bibinfo {pages}
  {1639} (\bibinfo {year} {1974})}\BibitemShut {NoStop}%
\bibitem [{\citenamefont {Turcs\'anyi}\ \emph {et~al.}(1988)\citenamefont
  {Turcs\'anyi}, \citenamefont {Puk\'anszky},\ and\ \citenamefont
  {T\"udos}}]{Turcsanyi-JMSL88}%
  \BibitemOpen
  \bibfield  {author} {\bibinfo {author} {\bibfnamefont {B.}~\bibnamefont
  {Turcs\'anyi}}, \bibinfo {author} {\bibfnamefont {B.}~\bibnamefont
  {Puk\'anszky}}, \ and\ \bibinfo {author} {\bibfnamefont {F.}~\bibnamefont
  {T\"udos}},\ }\href@noop {} {\bibfield  {journal} {\bibinfo  {journal} {J.
  Mater. Sci. Lett.}\ }\textbf {\bibinfo {volume} {7}},\ \bibinfo {pages} {160}
  (\bibinfo {year} {1988})}\BibitemShut {NoStop}%
\bibitem [{\citenamefont {Chen}\ and\ \citenamefont {Lin}(2005)}]{Chen-JMS05}%
  \BibitemOpen
  \bibfield  {author} {\bibinfo {author} {\bibfnamefont {J.-S.}\ \bibnamefont
  {Chen}}\ and\ \bibinfo {author} {\bibfnamefont {K.-Y.}\ \bibnamefont {Lin}},\
  }\href@noop {} {\bibfield  {journal} {\bibinfo  {journal} {J. Mater. Sci.}\
  }\textbf {\bibinfo {volume} {40}},\ \bibinfo {pages} {87} (\bibinfo {year}
  {2005})}\BibitemShut {NoStop}%
\bibitem [{\citenamefont {Bak}\ and\ \citenamefont
  {Sneppen}(1993)}]{BakSneppen-PRL93}%
  \BibitemOpen
  \bibfield  {author} {\bibinfo {author} {\bibfnamefont {P.}~\bibnamefont
  {Bak}}\ and\ \bibinfo {author} {\bibfnamefont {K.}~\bibnamefont {Sneppen}},\
  }\href@noop {} {\bibfield  {journal} {\bibinfo  {journal} {Phys. Rev. Lett.}\
  }\textbf {\bibinfo {volume} {71}},\ \bibinfo {pages} {4083} (\bibinfo {year}
  {1993})}\BibitemShut {NoStop}%
\bibitem [{\citenamefont {Schmittbuhl}\ \emph {et~al.}(1995)\citenamefont
  {Schmittbuhl}, \citenamefont {Roux}, \citenamefont {Vilotte},\ and\
  \citenamefont {M{\aa}l{\o}y}}]{Schmittbuhl-PRL95}%
  \BibitemOpen
  \bibfield  {author} {\bibinfo {author} {\bibfnamefont {J.}~\bibnamefont
  {Schmittbuhl}}, \bibinfo {author} {\bibfnamefont {S.}~\bibnamefont {Roux}},
  \bibinfo {author} {\bibfnamefont {J.~P.}\ \bibnamefont {Vilotte}}, \ and\
  \bibinfo {author} {\bibfnamefont {K.~J.}\ \bibnamefont {M{\aa}l{\o}y}},\
  }\href@noop {} {\bibfield  {journal} {\bibinfo  {journal} {Phys. Rev. Lett.}\
  }\textbf {\bibinfo {volume} {74}},\ \bibinfo {pages} {1787} (\bibinfo {year}
  {1995})}\BibitemShut {NoStop}%
\bibitem [{\citenamefont {Gao}\ and\ \citenamefont
  {Rice}(1989)}]{GaoRice-JAM89}%
  \BibitemOpen
  \bibfield  {author} {\bibinfo {author} {\bibfnamefont {H.~J.}\ \bibnamefont
  {Gao}}\ and\ \bibinfo {author} {\bibfnamefont {J.~R.}\ \bibnamefont {Rice}},\
  }\href@noop {} {\bibfield  {journal} {\bibinfo  {journal} {J. Appl. Mech.}\
  }\textbf {\bibinfo {volume} {56}},\ \bibinfo {pages} {828} (\bibinfo {year}
  {1989})}\BibitemShut {NoStop}%
\bibitem [{\citenamefont {Arratia}\ and\ \citenamefont
  {Gordon}(1989)}]{Arratia-BMB89}%
  \BibitemOpen
  \bibfield  {author} {\bibinfo {author} {\bibfnamefont {R.}~\bibnamefont
  {Arratia}}\ and\ \bibinfo {author} {\bibfnamefont {L.}~\bibnamefont
  {Gordon}},\ }\href@noop {} {\bibfield  {journal} {\bibinfo  {journal} {Bull.
  Math. Biol.}\ }\textbf {\bibinfo {volume} {51}},\ \bibinfo {pages} {125}
  (\bibinfo {year} {1989})}\BibitemShut {NoStop}%
\bibitem [{\citenamefont {Blondeau}(2011)}]{Blondeau-PhD11}%
  \BibitemOpen
  \bibfield  {author} {\bibinfo {author} {\bibfnamefont {C.}~\bibnamefont
  {Blondeau}},\ }\emph {\bibinfo {title} {La cryptanalyse diff\'erentielle et
  ses g\'en\'eralisations}},\ \href@noop {} {Ph.D. thesis},\ \bibinfo  {school}
  {Univ. Pierre et Marie Curie}, \bibinfo {address} {Paris} (\bibinfo {year}
  {2011})\BibitemShut {NoStop}%
\bibitem [{\citenamefont {Blondeau}\ \emph {et~al.}(2011)\citenamefont
  {Blondeau}, \citenamefont {G\'erard},\ and\ \citenamefont
  {Tillich}}]{Blondeau-DCC11}%
  \BibitemOpen
  \bibfield  {author} {\bibinfo {author} {\bibfnamefont {C.}~\bibnamefont
  {Blondeau}}, \bibinfo {author} {\bibfnamefont {B.}~\bibnamefont {G\'erard}},
  \ and\ \bibinfo {author} {\bibfnamefont {J.-P.}\ \bibnamefont {Tillich}},\
  }\href@noop {} {\bibfield  {journal} {\bibinfo  {journal} {Des. Codes
  Crypt.}\ }\textbf {\bibinfo {volume} {59}},\ \bibinfo {pages} {3} (\bibinfo
  {year} {2011})}\BibitemShut {NoStop}%
\bibitem [{\citenamefont {Weiss}\ \emph {et~al.}(2014)\citenamefont {Weiss},
  \citenamefont {Girard}, \citenamefont {Gimbert}, \citenamefont {Amitrano},\
  and\ \citenamefont {Vandembroucq}}]{WGGAV-PNAS14}%
  \BibitemOpen
  \bibfield  {author} {\bibinfo {author} {\bibfnamefont {J.}~\bibnamefont
  {Weiss}}, \bibinfo {author} {\bibfnamefont {L.}~\bibnamefont {Girard}},
  \bibinfo {author} {\bibfnamefont {F.}~\bibnamefont {Gimbert}}, \bibinfo
  {author} {\bibfnamefont {D.}~\bibnamefont {Amitrano}}, \ and\ \bibinfo
  {author} {\bibfnamefont {D.}~\bibnamefont {Vandembroucq}},\ }\href@noop {}
  {\bibfield  {journal} {\bibinfo  {journal} {Proc. Nat. Acad. Sci.}\ }\textbf
  {\bibinfo {volume} {111}},\ \bibinfo {pages} {6231} (\bibinfo {year}
  {2014})}\BibitemShut {NoStop}%
\end{thebibliography}
\end{document}